\documentclass{ieeeaccess}
\usepackage{cite}
\usepackage{amsmath,amssymb,amsfonts}
\usepackage{algorithmic}
\usepackage{graphicx}
\usepackage{textcomp}
\usepackage{subfigure}
\usepackage{comment}
\usepackage{soul}
\usepackage{booktabs}
\usepackage{tablefootnote}
\def\BibTeX{{\rm B\kern-.05em{\sc i\kern-.025em b}\kern-.08em
    T\kern-.1667em\lower.7ex\hbox{E}\kern-.125emX}}
\usepackage{comment}
\usepackage{setspace}

\begin{document}
\history{
%Date of publication xxxx 00, 0000, %date of current version xxxx 00, 0000.
}
\doi{
%10.1109/ACCESS.2021.DOI
}

\title{\mbox{Lite-Sparse} Hierarchical Partial Power Processing for Second-Use Battery Energy Storage Systems}
\author{\uppercase{Xiaofan Cui}\authorrefmark{1}, \IEEEmembership{Student Member, IEEE},
\uppercase{Alireza Ramyar}\authorrefmark{1}, \IEEEmembership{Student Member, IEEE},
\uppercase{Peyman Mohtat}\authorrefmark{2}, \IEEEmembership{Student Member, IEEE},
\uppercase{Veronica Contreras}\authorrefmark{1}, \IEEEmembership{Student Member, IEEE},
\uppercase{Jason B. Siegel}\authorrefmark{2}, 
\IEEEmembership{Senior Member, IEEE},
\uppercase{Anna G. Stefanopoulou}\authorrefmark{2}, \IEEEmembership{Fellow, IEEE}, and \uppercase{Al-Thaddeus Avestruz}\authorrefmark{1},
\IEEEmembership{Senior Member, IEEE}}
\address[1]{Electrical and Computer Engineering, University of Michigan, Ann Arbor, MI 48105 USA}
% \address[2]{Electrical and Computer Engineering, University of Michigan, Ann Arbor, MI 48105 USA (e-mail: aramyar@umich.edu)}
\address[2]{Mechanical Engineering, University of Michigan, Ann Arbor, MI 48105 USA}
\tfootnote{``This work was supported in part by the Michigan Transportation Research and Commercialization (MTRAC) Grant CASE-283536 of the 21st Century Jobs Trust Fund received through the Michigan Strategic Fund (MSF) from the State of Michigan. The work was also supported in part by the National Science Foundation under CAREER Award No. 2146490.''}

%\markboth
%{X. Cui et al. \headeretal: Optimizing Partial Power Processing Second-Use Battery Energy Storage Systems}{}
% {Author \headeretal: Preparation of Papers for IEEE TRANSACTIONS and JOURNALS}

\corresp{Corresponding author: Xiaofan Cui (e-mail: cuixf@umich.edu).}

\begin{abstract}
The growth of electric vehicles (EVs) will be followed by a surge in retired EV batteries, which could be repurposed since they might be having nearly 80\% available capacity. 
%One sustainable solution is to assemble these in 
Repurposing automotive batteries for second-use battery energy storage systems (2-BESS) has both economical and environmental benefits. 
The challenge with assembling and aggregating second-use batteries to work together in a system is the heterogeneity in their capacity and power limits that can be evolving based on their degrading state of health.
This paper introduces a new strategy to optimize 2-BESS performance despite the heterogeneity of individual batteries while reducing the cost of power conversion. In this paper, the statistical distribution of the power heterogeneity in the supply of batteries is used to optimize the choice of power converters and design the power flow within the battery energy storage system (BESS) to optimize power capability. By leveraging a new \mbox{lite-sparse} hierarchical partial power processing (LS\nobreakdash-HiPPP) approach, we study how a hierarchy in partial power processing (PPP) partitions power converters to significantly reduce converter ratings, process less power to achieve high system efficiency with lower cost (lower efficiency) converters, and take advantage of economies of scale by requiring only a minimal number of sets of identical converters. Our results demonstrate that LS\nobreakdash-HiPPP architectures offer the best tradeoff between battery utilization and converter cost and have higher system efficiency than conventional partial power processing (C-PPP) in all cases.
\end{abstract}
\begin{keywords}
Battery energy storage systems (BESS), Second-use battery energy storage systems (2-BESS),  second-use batteries, second-life batteries, repurposed batteries

% Enter key words or phrases in alphabetical 
% order, separated by commas. For a list of suggested keywords, send a blank 
% e-mail to keywords@ieee.org or visit \underline
% {http://www.ieee.org/organizations/pubs/ani\_prod/keywrd98.txt}
\end{keywords}
\titlepgskip=-15pt
\maketitle

\section{Introduction}
\label{sec:introduction}
\PARstart{B}atteries from retired electric vehicles (EVs) represent both a problem and an opportunity. By 2030, there will be 200 GWh per year of used batteries from EVs \cite{engel2019}, that can be used in stationary and less demanding applications to maximize their utilization before recycling them.
These batteries, when removed from the vehicle, still have approximately 80\% capacity and power capability when compared to the fresh. Reusing these batteries in second-use battery energy storage systems (2-BESS) reduces mining of raw material and manufacturing costs and environmental burdens. Moreover, they add economic value to EV batteries.

There are several economic obstacles to the adoption and deployment of 2\nobreakdash-BESS. The price competitiveness of 2\nobreakdash-BESS relative to 
%other storage technologies including battery energy storage system (BESS) with 
new batteries relies on lowering the added costs. These include the cost of transportation \cite{slattery2021transportation}, inventory, and power converters \cite{Neubauer2015}. 
%By using distributed \red{production} and a local supply together with just-in-time \red{production}, transportation and inventory costs are minimized. 
By distributing the assembly of 2-BESS geographically, one can use the local supply together with just-in-time assembly \cite{Pinto2018} to gather the production, transportation, and inventory cost. However, this strategy incurs the challenge of a heterogeneous supply.
Even with second-use batteries that are identical at the time of original manufacturing and installed in identical vehicles, these batteries, when removed, will exhibit a significant degree of variation because of the different history of drive cycles and temperature cycling.

BESSs are needed to stabilize the grid with a high penetration of renewables \cite{Neto2018}, support micro- and nano-grids \cite{Bryden2019}, support EV fast charging, and reduce the cost of grid upgrades from the high peak power \cite{DArpino2019}. A canonical network for a BESS is illustrated in Fig.\,\ref{fig:3NetComp}(a), which is a network of batteries interconnected with a network of power converters that determine the power flows, and an output to a load.

% \Figure[t!](topskip=0pt, botskip=0pt, midskip=0pt){"BESS_network.jpg"}
% {Battery Energy Storage Network.
% Batteries are interconnected with bi-directional power converters.\label{fig_c-ppp}}

% \Figure[t!](topskip=0pt, botskip=0pt, midskip=0pt)[width = 16cm]{"3NetComp.png"}
% {(a) Battery Energy Storage Network. Batteries are interconnected with bi-directional power converters; (b) Full Power Processing. A power converter that is rated at the full power capability of the battery is needed for each battery, resulting 100 \% of the output power being processed; (c) Conventional Partial Power Processing. Partial-power processing for series-connected batteries requires one power converter for every battery. Only the mismatch power is processed by the power converters. \label{fig:3NetComp}}

% \begin{figure*}[ht]
%     \centering
%     \includegraphics[width= 16 cm]{./Graphics/3NetComp.png}
%     \caption{(a) Battery Energy Storage Network. Batteries are interconnected with bi-directional power converters; (b) Full Power Processing. A power converter that is rated at the full power capability of the battery is needed for each battery, resulting 100 \% of the output power being processed; (c) Conventional Partial Power Processing. Partial-power processing for series-connected batteries requires one power converter for every battery. Only the mismatch power is processed by the power converters.}\label{fig:3NetComp}
% \end{figure*}

\Figure[t!](topskip=0pt, botskip=0pt, midskip=0pt)[width=\linewidth]{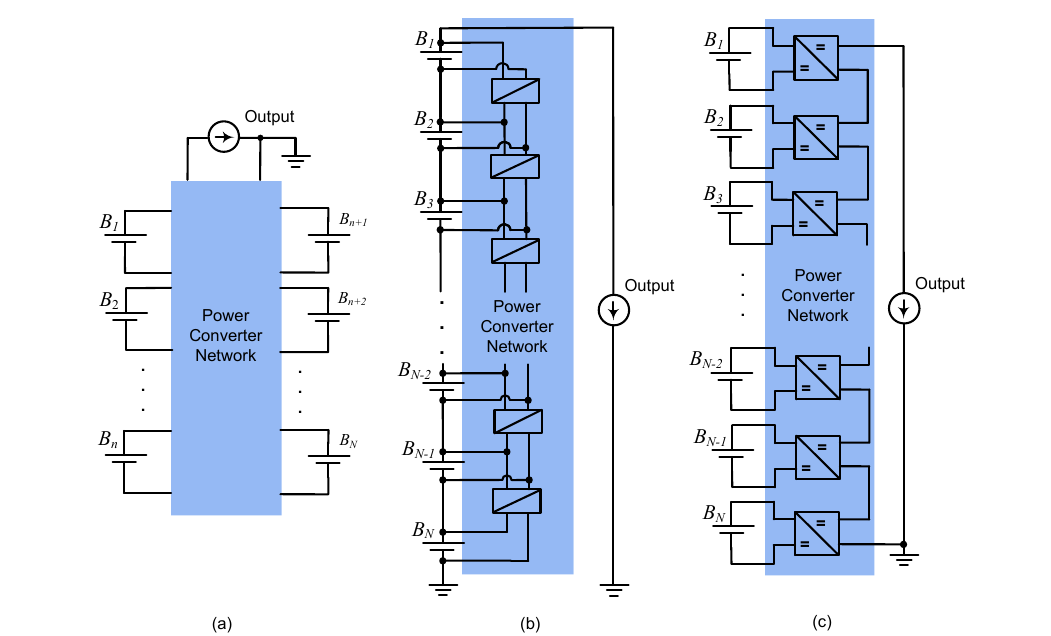}{(a) Battery Energy Storage System (BESS) consists of batteries connected to a network of power converters.\newline  (b) Conventional Partial Power Processing (C-PPP) in this particular topology cascades power from neighboring batteries and processes only the mismatch power.
%The bus voltage regulator processes only the mismatch between the batteries in series and the required output voltage. 
(c) Full Power Processing (FPP) uses a power converter for each battery to individually determine the charge and discharge current, voltage, and hence power trajectories. FPP delivers in this series case individually specified voltages to the output load. \label{fig:3NetComp}} 

%\Figure[t!][width = 16cm ]{"3NetComp.png"}{(a) Second-Use Battery Energy Storage System (2-BESS) consists of second-use (2U) batteries connected to a network of power converters.\newline (b) Full Power Processing (FPP) uses a power converter for each battery to individually determine the charge and discharge current, voltage, and hence power trajectories. FPP delivers in this series case individually specified voltages to the output load.  (c) Conventional Partial Power Processing (C-PPP) processes only the mismatch power} \label{fig:3NetComp}

A typical strategy for BESSs using new batteries, which have a high degree of homogeneity, is to use conventional partial power processing (C-PPP) architectures\cite{Shenoy2013a}. Partial power processing reduces the required power converter ratings\footnote{\emph{Power rating} is a quantity that describes the total electrical power required for normal operation of a power converter.}\cite{Kim2015b} and hence the capital cost of converters \cite{Shenoy2013a}. Additionally, by reducing processed power, overall system efficiency increases and the cost of thermal management decreases \cite{Candan2015}. These architectures can have a pre-determined choice of power converters and power flow topology because of both the high certainty in and homogeneity among the batteries \cite{Evzelman2016} they process. Fig.\,\ref{fig:3NetComp}(b) shows a typical partial power processing topology for batteries in series. Only the mismatched power, which is relatively small, rather than the full power is processed.
This is true for power capability mismatch among batteries as well as the voltage mismatch between the entire BESS and the load.

% \Figure[t!](topskip=0pt, botskip=0pt, midskip=0pt)[width= 6 cm]{"fppp_inNet.png"}
% {Full Power Processing.  A power converter that is rated at the full power capability of the battery is needed for each battery, resulting 100\% of the output power being processed.\label{fig_full-pp}}

% \Figure[t!](topskip=0pt, botskip=0pt, midskip=0pt) [width= 6 cm] {"cppp_inNet.png"}
% {Conventional Partial Power Processing.
% Partial-power processing for series-connected batteries requires one power converter for every battery. Only the mismatch power is processed by the power converters.\label{fig_bess_network}}

The conventional strategy for BESSs with heterogeneous batteries, e.g. 2-BESS, is to individually process all the power from every battery to adjust for the heterogeneity by individualizing each battery's power trajectory \cite{Mukherjee2015}. Fig.\,\ref{fig:3NetComp}(c) illustrates the full power processing (FPP) strategy; the disadvantage to this approach is that the power ratings of the converters must be at least equal to the battery power ratings. Because power converter cost is typically nearly proportional to their power rating, FPP is the costliest option for power conversion. Additionally, the system efficiency will be less than the efficiency of the power converters \cite{Rancilio2019}. For example, a system efficiency of 98\% requires power converters with at least 98\% efficiency, where the cost of the converters also increases with efficiency \cite{Biela2011}. 

An appealing alternative to FPP for 2\nobreakdash-BESS is partial power processing because of the lower cost of power converters, higher system efficiency, and lower cooling requirements.  However, heterogeneity among batteries is a challenge. This paper demonstrates the disadvantages of FPP and the challenges of C-PPP approaches in comparison to a new strategy for partial power processing, \mbox{Lite-Sparse} Hierarchical Partial Power Processing (LS\nobreakdash-HiPPP). This paper presents the formulation and results for the optimization of power capability in LS\nobreakdash-HiPPP.
%\!\footnote{An energy optimization formulation is also applicable: for example, when optimizing the energy buffering needed for supporting the grid in EV fast charging over a time trajectory. \red{The following was submitted to another journal and is available as an arXiv preprint \cite{cui2021grid}.}}

As the heterogeneity of batteries grows, C-PPP and FPP are even more inefficient.
The strategy presented in this paper for optimizing a 2\nobreakdash-BESS uses LS\nobreakdash-HiPPP to accommodate the heterogeneity in the power capability of the individual batteries.  With LS\nobreakdash-HiPPP, we show that the overall output power capability is not heavily compromised as it is with conventional power processing architectures using the same converter ratings.  In addition, the overall power capability of a 2-BESS with LS-HiPPP is less sensitive to battery heterogeneity, which results in a higher power derating;\footnote{\emph{Power derating} means the percentage of the rating specification that is guaranteed to high certainty. More details about power derating can be found in Section\,\ref{sec:power_derating}.} together with higher power capability, this is more cost effective, which is represented by a higher power captured value as discussed in Section\,\ref{sec:results}.

% State-of-the-art modeling techniques of the batteries that have been used in EV applications include} \cite{Smith2012, Sarasketa-Zabala2014, Cordoba-Arenas2015a, Samad2018}.
% \cite{Smith2012, Sarasketa-Zabala2014} model the effects of geographies, driving cycles, and storage conditions on energy capacity degradation of batteries. The power capability fading of batteries is fitted by a semi-empirical cycle-life model} \cite{Cordoba-Arenas2015a}.
% \cite{Samad2018} models and optimizes the sizing of EV battery packs to improve the energy opportunities. For general batteries, the modeling of the other characteristics such as thermoelectric interaction} \cite{Liu2017a}, thickness expansion} \cite{Mohtat2021}, and differential voltage} \cite{Chen2022} have been investigated.} \cite{Liu2017a} captures the coupled thermoelectric behaviors of batteries by an auto-regressive moving average model.
% The state of health of batteries is correlated with the thickness expansion model in }\cite{Mohtat2021} and differential voltage model in }\cite{Chen2022}.

Many battery modeling efforts have focused on describing the State of Health (SoH), namely capacity loss and internal resistance growth \cite{Edge2021} which is particularly important for managing second-use batteries. The battery SoH can be extracted from low-rate cycling data by considering the shift in peaks of the differential voltage \cite{Chen2022, Lee2020}. Empirical models have been widely-used to describe the energy capacity degradation and power capability fading of batteries over time and usage based on the operating conditions \cite{Jin2018, Cordoba-Arenas2015a}, however the empirical models require a significant data collection effort for training so that the models are generally applicable for predicting degradation over different usage cases \cite{Sulzer2021}.

State-of-the-art lithium-ion battery models based on the Dual-Foil Neman type Pseudo-2D model can be simplified into single particle models \cite{Marquis2019} for grid scale energy storage applications with relatively low C-rates corresponding to multi-hour discharge requirements. These models are able to capture the impact of operating conditions, temperature, state of charge, and charging currents on the rates of capacity loss and resistance growth \cite{Edge2021}. The relevant degradation mechanism included in the physics-based models include lithium plating \cite{Yang2017a}, stress and particle cracking which leads to loss of active material \cite{Ai2020a, Christensen2006}, and growth of the solid electrolyte interface (SEI) layer \cite{Deshpande2017}.

Semi-empirical models can also be effective for describing degradation \cite{Varini2019}. The coupling of battery performance, thermoelectric behavior and degradation can also be addressed by an auto-regressive moving average model \cite{Liu2017a}. In the future, the energy management strategy for 2-BESS could perform the wear leveling of battery assets by considering the coupled thermal and degradation mechanisms to extend their life.

Batteries which are close to, or well beyond the knee point in their capacity loss may be unsuitable for use in 2BESS, however identifying and classifying the battery state of health is challenging \cite{Kwon2020}. Machine learning techniques can be used to help identify suitability of battery assets for use in 2BESS and optimum replacement schedules based on where the battery capacity lies with respect to the knee point in capacity degradation rate \cite{Fermin-Cueto2020a}. In addition, the machine learning techniques could be applied directly to parameters of the physics-based models to improve the prediction of future degradation \cite{Liu2022, Liu2022a, jiang2020}.

State-of-the-art models for the BESS power processing include the FPP, C-PPP, and hierarchical PPP.
FPP include the series \cite{Mukherjee2015} and parallel \cite{moo2008parallel} architectures.
C-PPP architectures are different in the graphical patterns of the battery energy exchange. \cite{Kim2014a, Hua2016, Ye2015, Stauth2013} only allow the adjacent two batteries to exchange the energy. \cite{Einhorn2011, Mestrallet2014} allow each battery to partially deliver the power to the output bus. \cite{Evzelman2016} substitutes this output bus with an extra dc bus that is formed by a capacitor. In comparison, \cite{Lim2014a, With2015a, Li2013b} use an extra ac bus for the energy exchange.
A few types of hierarchical PPP architectures \cite{Dong2015b, Chen2015b, Zhang2017a} have appeared in the literature as the SoC balancers for new batteries.
%alternative approach to do the statistical design of BESS is to use AI and machine learning [1][2].

The specific list of contributions is as follows:
\begin{enumerate}
\item A new lite-sparse hierarchical partial power processing (LS-HiPPP) architecture is introduced in Section\,\ref{sec:methods}.
The LS-HiPPP achieves the best 2-BESS performance compared to the state-of-the-art power processing architectures despite the large heterogeneity of individual batteries.
The advantages of LS-HiPPP over the existing state-of-the-art architectures C-PPP and FPP are compared in Table\,\ref{table:pp_compare}.
\begin{table}[tbp]
    \caption{Comparison of Three Power Processing  Architectures \tablefootnote{Given the conditions of 85\% converter power efficiency and 20\% battery power heterogeneity. \emph{Power heterogeneity} is represented by the standard deviation $\hat{\sigma}_p$ normalized to the expected power. More details about power heterogeneity can be found in Section\,\ref{sec:methods}.}}
    \label{table:pp_compare}
    \centering
    \begin{tabular}{cccc}
    \toprule
        \textbf{Architecture} &\textbf{FPP} &\textbf{C-PPP} &\textbf{LS-HiPPP} \\
        \textbf{} &\textbf{} &\textbf{} &(This Paper) \\
        \midrule
        \textbf{Efficiency}\tablefootnote{\emph{Efficiency} refers to the electrical system efficiency and does not include the auxiliary power consumption \cite{Rancilio2020}, such as cooling and control.} & 85\,\% & 91\,\% & 97\,\%  \\
        \midrule
        \textbf{Utilization} & 100\,\% & 81\,\% & 95\,\%  \\
        \midrule
        \textbf{Relative Cost}\tablefootnote{\emph{Cost} refers to the normalized cost of power converters for constant \$/kW. The base value is the power converter cost of LS-HiPPP and C-PPP.} & 5\,$\times$ & 1\,$\times$ & 1\,$\times$ \\
    \bottomrule
    \end{tabular}
\end{table}
\item Optimization of LS-HiPPP for maximizing the power output is explained in Section\,\ref{sec:methods}. A new Distribution Flattening method enables the power processing architecture optimization.
\item Demonstration of significant improvement for 2-BESS using LS\nobreakdash-HiPPP in battery power utilization, electrical system efficiency, power derating, and power captured value, as shown in Section\,\ref{sec:results}.
\end{enumerate}

% The paper is organized as follows:  Section\, \ref{sec:methods} discusses the \mbox{Lite-Sparse} Hierarchical Partial Power Processing (LS\nobreakdash-HiPPP) architecture and optimization strategy, and Section III presents the simulation results of the performance comparisons between LS\nobreakdash-HiPPP and conventional power processing strategies.

\section{Methods and Theory}\label{sec:methods}
An objective often used for optimization in operations research over statistical uncertainties is the ensemble performance in the production of a large number of units, specifically the expected performance \cite{Arrow1957}.  Expected performance metrics are used for optimization and evaluation in the subsequent sections of this paper.

The design targets for power processing in a 2-BESS are: (1) minimize the aggregate power rating of the power conversion; (2) minimize the number of power converters; (3) minimize the different types of converters; and (4) maximize the overall performance, specifically power capability in this paper.  A goal of this research is to find the optimal tradeoff surface for (1) and (4) using optimization methods, with (2) and (3) as design choices that depend on the pricing structure of power converters.  The choice of design point on the (1) and (4) tradeoff surface depends on the pricing structures of both batteries and power converters.
\begin{comment}
Talk about metrics & specific (for this paper) objectives using those metrics
\end{comment}

The {\em battery utilization} is the fraction that is available at the output of the combined individual capabilities of the batteries within a BESS.  In this paper, the expected battery power utilization is maximized given the statistics of the battery supply and choice of power converters. This more convenient optimization is a type of duality to minimizing the power processing for a given choice of battery utilization.

The aggregate power converter rating of a BESS is the sum of the ratings of the individual power converters within the energy storage system.  The cost of power converters is known to be monotonic with power rating with fewer types of converters being more advantageous for economies of scale.

System efficiency is the ratio of the output power of a BESS to the sum of the power delivered by the individual batteries.  For full power processing, 100\% of the individual power is processed by the power converters, which means system efficiency is determined by power converter efficiency. Section\,\ref{sec:sys_eff} discusses how partial power processing increases system efficiency without requiring more efficient power converters. More efficient power converters are typically larger in size and more expensive. Lower system efficiency means higher losses, which means a larger cost in thermal management.

Heterogeneity in the battery supply creates performance variation in the 2-BESS.  In this paper, we address the sensitivity of output power to heterogeneity.  This can be derived from the Monte Carlo results of battery power utilization in Section\,\ref{sec:results}. The statistics of the output performance of the 2-BESS determine both the derating and the captured value.

The {\it power derating} is the power capability of a particular 2-BESS within some confidence level when the batteries are drawn from a statistically distributed supply, which is related to the derating in \cite{Cui2022}.  The {\it power captured value} is a proxy for the expected financial value of a 2-BESS capability; in the context of this paper's analysis, it is the product of power derating and power utilization \cite{Cui2022}.
High power derating and captured value imply a high-performance energy storage architecture.

\subsection{2-BESS Architecture: \mbox{Lite-Sparse} Hierarchical Partial Power Processing}
The approach to partial power processing in this paper is circuit interconnection that is hierarchical, where most of the converters are ``lite'' in power with a sparse number of converters with more power.  This results in a much lower processed power and hence aggregate converter rating for a particular 2-BESS performance.  We use the hierarchy in the partial power processing to partition the power converters to take advantage of economies of scale by requiring only a minimal number of sets of identical power converters.  This way, only a few types of power converters are needed, which can be purchased in larger volumes.  The combination of numerous lower power converters together with a few higher power converters comprise the power processing for LS\nobreakdash-HiPPP.

The optimization of LS-HiPPP is constrained by the statistical distribution of the battery supply.  This paper will illustrate the strategy for a familiar and frequently used series interconnection of batteries, within which the power converter choice, interconnection, and power flow will be designed, as illustrated in Fig.\,\ref{fig:LS-HiPPP2}.

\Figure[t!](topskip=0pt, botskip=0pt, midskip=0pt)[width = 8.5cm]{LS-HiPPPWBVR.pdf}
{\mbox{Lite-Sparse} Hierarchical Partial-Power Processing (\mbox{LS-HiPPP}) for series connected 2-BESS.  Layer 1 consists of a sparse set of higher power converters.  Layer 2 consists of a dense set of lower power (lite) converters.  A bus voltage regulator processes the mismatch between the battery series string and the required bus voltage.  Only mismatch power is processed like C-PPP but with fewer power converters and lower converter ratings for the same performance using heterogeneous second-use batteries.\label{fig:LS-HiPPP2}}

In contrast to FPP where every battery requires its own power converter to process its power, the LS\nobreakdash-HiPPP interconnection in this paper consists of two power converter layers. The first layer consists of a sparse number of power-heavy converters that is much fewer in number than the batteries.  The second layer consists of a second more dense layer of lite-power converters.

The {\it Design} of a battery storage network is the instantiation of architecture and can be defined by the following:
\begin{enumerate}
    \item Set of batteries whose elements possess the relevant characteristics, for example, power capability. Other relevant characteristics can include the statistical distribution of supply, which can be parameterized by their moments, e.g. mean and variance.
    \item Interconnection of batteries, which can be represented by a graph or a circuit.
    \item Power processing Design.
\end{enumerate}

The BESS shown in Fig.\ref{fig:LS-HiPPP2} with the Layer 1 Sparse and Layer 2 Lite converters includes a {\em Bus Voltage Regulator}. The Lite-Sparse converters only process the mismatch power among the series connected batteries, but unlike conventional power processing, requires lower overall power converter ratings.  The Bus Voltage Regulator processes the voltage mismatch between the series string of batteries and the voltage required by the current-sink load; in doing so, the voltage heterogeneity is also absorbed. A parallel LS-HiPPP implementation is also possible and is discussed in \cite{Cui2021c}.

The methods discussed in this paper apply not only to battery packs but also to battery modules (battery packs may be partitioned into modules) and individual battery cells.

\subsection{Bus Voltage Regulator}
The battery string voltage and the desired output voltage might be different as the batteries are discharged. The bus voltage regulator is designed to compensate for this voltage mismatch dynamically. The bus voltage regulator recycles some of the output power of the 2-BESS to support the voltage mismatch \cite{Zhou2015}, which can be positive or negative, between the battery string and load. The output voltage of the 2-BESS is regulated using feedback. Although the system efficiency is reduced, the bus voltage regulator uses partial power processing to minimize the impact \cite{Zhou2015}. For example, an 85\% efficient bus voltage regulator supporting a 10\% voltage mismatch will reduce the system efficiency by 1.5\%.

\subsection{Model Specifics}

\subsubsection{Battery Macromodel}
A Gaussian distribution is used for the power capability of the supply of batteries.  Although the methods presented in this paper can be used for any distribution, including both discrete and continuous, Gaussian distributions are typically used in analysis and comparisons among technology or configuration options.

\subsubsection{Model Mapping Using Distribution Flattening}
%Model for Statistics of Individual Battery Positions -- Distribution Flattening
The battery supply statistics can be mapped to the statistics of individual battery positions by a Distribution Flattening method.
% Macro-model --> Micro-Model
% Model mapping
The method of Distribution Flattening generates a finite set of batteries that represent the expected performance.  Fig.\,\ref{fig:flattening} shows a probability density function (PDF) for battery power capability $p(P)$.

% \noindent The following is directly quoted and equations modified from \cite{cui2021grid}:
% \par
% \begingroup
% \leftskip4em
% \rightskip\leftskip 
We would like to map this statistical distribution, which is a continuous function, to a finite {\it expected set} of batteries of size $N$.  In this paper, the statistical distribution of the supply is Gaussian, but this does not necessarily have to be the case.  

This {\it expected set} is an ordered set.  The elements of this set are a particular representation of the expected values for $N$ batteries drawn from the supply distribution.  The set is constructed in the following manner:
\begin{enumerate}
    \item Divide the distribution into $N$ intervals of equal probability: $[P_1,P_2]$, $[P_2,P_3]$, \ldots, $[P_N,P_{N+1}]$. $P_1$ and $P_{N+1}$ are the lower and upper bounds of battery power capability, respectively. An example is shown in Fig.\,\ref{fig:flattening}. The $n^{\text{th}}$ interval satisfies 
    \begin{align}
        \int_{P_n}^{P_{n+1}} p(P)\,dP = \frac{1}{N}.
    \end{align}
    \item Assign each interval its expected value (1st moment).
         {\begin{equation}
             \bar{P}_n =  N\int_{P_n}^{P_{n+1}} p(P)P\,\text{d}P.
         \end{equation}}
    \item The finite expected set $\mathcal{B}$ is constructed as $\mathcal{B} = \{\bar{P}_1, \bar{P}_2, \ldots, \bar{P}_N\}$ .
\end{enumerate}
%\par
%\endgroup
In general, each interval can be assigned any measure of central tendency, including those that are functions of the local shape of the interval.  For example, one could use a function of the higher moments of the interval.
Fig.\,\ref{fig:flattening} shows a realization of $\mathcal{B}$ as a series circuit of batteries.  In general, $\mathcal{B}$ can be realized by any topology, including circuit topologies.

\subsubsection{Component Modeling}
The batteries used in practice have a high efficiency so they can be modeled to have negligible loss. A similar argument can be made that the optimization results using this approximation will deviate by only a small amount from the true optimum.

Because each converter processes only a fraction of the power that is extracted from the batteries, the optimization result using converters with negligible loss is expected to have only a small deviation from the true optimum.
%{\color{blue}
% \subsubsection{Operational Model} 
% The specific energy storage interconnection for LS\nobreakdash-HiPPP that we examine is illustrated in Fig.\,\ref{fig:LS-HiPPP2}. The topology specifically presented in this paper is a series string of batteries.
% The output load is a fixed current source.

% (I think we don't need this paragraph... The voltage mismatch between the battery stack and required bus voltage is regulated and processed by a power converter. The voltage heterogeneity among the batteries in a series stack results in an aggregate voltage mismatch between the battery stack and the required load voltage.
% The bus voltage regulator in Fig.\,\ref{fig:LS-HiPPP2} is a bidirectional converter that processes only the mismatched power; the power converter functionally in this case can be classified as partial power processing.  This bus voltage regulator can output either a positive or negative voltage to the bottom of the of the series stack and can be either non-isolated or isolated.)
%}

\Figure[t!](topskip=0pt, botskip=0pt, midskip=0pt)[width=\linewidth]{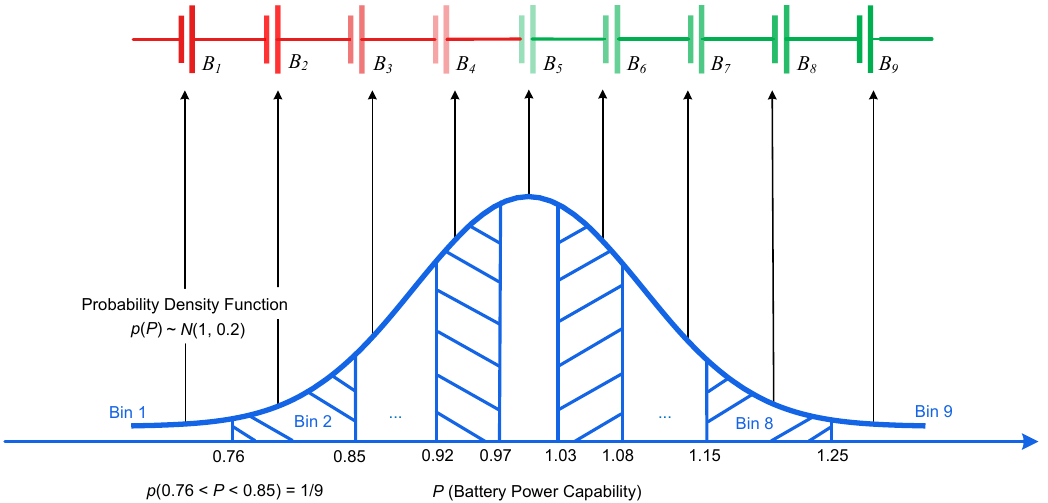}{Distribution Flattening Method maps a statistical distribution to a series string of batteries that represents the expected behavior for that string.  The design of the sparse Layer 1 converters uses this series string. \label{fig:flattening}}

\subsection{Optimizing Power Processing Design}
% {\color{blue} section C and D are reorganized, the outline is as following}
% % \red{start with talking about battery utilization , the objective of optimization}
% \begin{enumerate}
% {\color{blue}
%     \item Power processing Design activities: The first activity is to pre-determine the sets of power converters that will be used in the production of 2-BESS units. 
%     The second activity is the design and construction of a particular 2-BESS unit during actual production 
%     \item Optimization Objective — Battery Utilization
%     \item Optimization Constraints: Interconnections of input and output ports, Variables  of  the  input  and  output  ports,  e.g. voltage,current, and power.
%     \item Optimization Decision Variables: Sets of power converters, which can be parameterized by unique  set of power converter ratings  (e.g. power rating) and number of power converters in the set.
% Interconnection of power converters to the batteries.
%   Power flows among converters and batteries.
%   \item Optimization Formulation: Layer 1 power processing design, Layer 2 power processing design
%   }
% \end{enumerate}

The power processing design can be divided into two activities: (1) {\it allocation} and (2) {\it production}. The goal of the first activity is to pre-determine the sets of power converters that will be allocated for use in the production of 2-BESS units. The second activity is the design and construction of a particular \mbox{2-BESS} unit during actual production.

The power processing \mbox{design} of a battery storage network (in this paper a 2-BESS) can be defined by the following set, which comprises:
\begin{enumerate}
    \item Sets of power converters, which can be parameterized by a unique set of power converter ratings (e.g. power rating) and the number of power converters in the set.
    \item Interconnection of power converters to the batteries.
    \item Power flows among converters and batteries.
    \item Interconnections of input and output ports.
    \item Variables of the input and output ports, e.g. voltage, current, and power.
\end{enumerate}

In general, the power flows and the input and output port variables can be trajectories that change in continuous or discrete time, or in a sampled data space that may or may not be uniform in continuous-time intervals.

% In this paper, we specifically discuss the particular 2-BESS in a LS\nobreakdash-HiPPP architecture shown in Fig.\,\ref{fig:LS-HiPPP2}.  The circuit interconnection of the batteries is a series string.  The set of batteries is drawn from a supply population whose statistical distribution is a Gaussian parameterized by its mean and variance.  

The LS\nobreakdash-HiPPP structure that we are investigating consists of two layers of bidirectional power converters.  Layer 1 is a sparse layer of power converters that is optimized for the realization of an expected battery set from the supply population.  Layer 2 is a dense layer of power converters, i.e. the number of converters is equal to one fewer than the number of batteries, with each converter's ports attached to a battery and its adjoining neighbor, as shown in Fig.\,\ref{fig:LS-HiPPP2}.

\subsubsection{Decision Variables }\label{sec:converter-sets}
The power ratings of both Layer 1 and Layer 2 converters need to be determined.  As previously discussed, the structure we choose for LS\nobreakdash-HiPPP in this paper has the interconnection of Layer 2 converters pre-determined.  We also stipulate that the Layer 2 converters will be identical. For Layer 1, the number of converters and how they are partitioned into sets of identical converter ratings are determined as part of the optimization.

The cost of power converters scales approximately linearly with power rating (i.e. \$/kW).  There is a penalty in the Design as the number of power converter sets increases.  In other words, as the number of different types of converters that are needed for the Design of a particular BESS product increases, the worse the economies of scale because fewer converters of a particular type are purchased. 

\subsubsection{Optimization Objective: Battery Utilization}
In this paper, the objective function is the power utilization of the 2-BESS.  The power utilization $\mathcal{U_P}$  is defined as the total power delivered from the output port of the 2-BESS normalized by the sum of the intrinsic power capability of each individual battery in the 2-BESS.
\begin{align}
    \mathcal{U_P} \triangleq \frac{P_{\text{out}}}{\sum\limits_{j =1}^{N} P_j^b}, 
\end{align}
where $P_j^b$ represents the intrinsic power of the $j^{\text{th}}$ battery in the 2-BESS.

For simplicity, we assume that there is no penalty for power processing in the bus regulation converter.  This assumption is homologous to restricting battery sets to contain only batteries with identical voltages.  This does not mean that this design method is restricted only to batteries of identical voltages, but rather, the voltage differences do not change the optimization.

% The battery power utilization $\mathcal{U_P}$ is the output power capability $P_{\text{out}}$ of the 2-BESS normalized by the aggregate intrinsic battery power of a particular 2-BESS instantiation

\subsubsection{Optimization Formulation}
\paragraph{Layer 1 Power Processing Design}\label{sec:layer1design}
The purpose of Layer 1 is to process the expected mismatch power among heterogeneous batteries. 
The design flow of the power ratings of Layer 1 converters is illustrated in Fig.\,\ref{fig:layer1_production}.

The power processing for Layer 1 is designed from the expected set $\mathcal{B}$ that is derived from the Distribution Flattening of the battery supply set.  The series string of batteries $B_1 \ldots B_9$ are arranged from lowest to highest expected power capability as shown in Fig.\,\ref{fig:flattening}.

A mixed-integer optimization is performed to maximize the power utilization.  For a small number of converters, an exhaustive search can be performed to find the best interconnection.  For every interconnection, the optimal power flow is found using linear programming
\begin{align}
\underset{p^{(1)}_i, \,p^{\text{bat}}_j,\, p^{\text{bus}}_j}{\mathrm{max}}& \,\,\, \sum\limits_{1\le j \le N} p_j^{\text{bus}} \label{eqn:opt_layer1_obj} \\
\text{subject to} & \,\,\,\ -\bar{P}_j \le p^{\text{bat}}_j \le \bar{P}_j \label{eqn:opt_layer1_contraint1},\\
& p^{\text{bat}}_j = \sum\limits_{i \in K^{(1)}_j} p^{(1)}_i + p_j^{\text{bus}}, \label{eqn:opt_layer1_contraint2}\\
& p_j^{\text{bus}} = I_{\text{string}}V_j^{\text{bat}}, \quad j = 1,\,2,\,\ldots,\,N,\label{eqn:opt_layer1_contraint3}
\end{align}
where $K^{(1)}_j$ is the index set of the Layer-1 converters whose inputs are connected to the $j^{\text{th}}$ battery, and $N$ is the number of batteries in the string. In this optimization problem, the decision variables are the power ratings of the $M$ Layer 1 converters (represented by $p^{(1)}_1 \ldots p^{(1)}_M$), the power delivered from the $j^{\text{th}}$ battery to the bus (represented by $p_j^{\text{bus}}$), and the output power of $j^{\text{th}}$ battery (represented by $p_j^{\text{bat}}$).
The optimization objective (\ref{eqn:opt_layer1_obj}) maximizes the total power transferred to the bus (i.e. the summation of the power exchanged between each battery and the bus). 
Constraint (\ref{eqn:opt_layer1_contraint1}) are the battery input and output power limits. 
Constraint (\ref{eqn:opt_layer1_contraint2}) is the power conservation law for each battery. Constraint (\ref{eqn:opt_layer1_contraint3}) indicates that the direct power transfer from the $j^{\text{th}}$ battery to the bus is proportional to its voltage $V_j^{\text{bat}}$ because all the batteries share the same string current $I_{\text{string}}$.

At the maximum utilization, the converters will have optimal interconnection ${K_j^{(1)}}^{*}$ and power flow ${p^{(1)}_1}^{*} \cdots {p^{(1)}_M}^{*}$.  In general, each converter can be processing a different power, which can be described as the {\it optimal set of processed power}.  The converter ratings for Layer 1 can be partitioned into sets based on the set of different processed power for each converter.  For example, for a partitioning that consists of a single converter set, the rating for all the converters will be the highest processed power from the optimal set; for two partitions, the converter with the highest processed power will be one partition and the remaining converters will be rated at the 2\textsuperscript{nd} highest processed power, and so forth for more partitions.  This partitioning strategy results in the lowest aggregate rating for the power converters and hence cost.  It is worth noting that for a sparse Layer 1, the number of partitions is much fewer than the number of batteries.

\paragraph{Layer 2 Power Processing Design}\label{sec:layer2design}
The purpose of Layer 2 is to process the mismatch power from the statistical variation of the batteries.  To optimize over statistical variations, Monte Carlo methods are employed.
The design of Layer 2 proceeds subsequent to the design of Layer 1.  The interconnection and power ratings of the Layer 1 converters become constraints in the design of Layer 2 as illustrated in Fig.\,\ref{fig:layer2}.  As previously mentioned, the interconnection of Layer 2 is pre-determined and ratings of the power converters are identical; hence, the goal is to determine the optimal ratings for the power converters.
 
The optimization starts by selecting a set of trial power converter ratings. For each converter rating, an optimal set of battery power utilizations is obtained via a set of samples from the statistical distribution of the battery supply.  These battery power utilizations are calculated from the optimal power flow by applying linear programming to each sample
\begin{align}
\underset{p^{(1)}_i,\,p^{(2)}_k,\,p^{\text{bat}}_j,\,p^{\text{bus}}_j}{\mathrm{max}}& \,\,\, \sum\limits_{1\le j \le N} p_j^{\text{bus}} \label{eqn:opt_layer2_obj} \\
\text{subject to} & \,\,\,\ - (\bar{P}_j+\delta P_j) \le p^{\text{bat}}_j \le (\bar{P}_j + \delta P_j) \label{eqn:opt_layer2_contraint1},\\
& p^{\text{bat}}_j = \sum\limits_{i \in {K^{(1)}_j}^{*}}  p^{(1)}_i + \sum\limits_{k \in K^{(2)}_j} p^{(2)}_k + p_j^{\text{bus}}, \label{eqn:opt_layer2_contraint2}\\
& p_j^{\text{bus}} = I_{\text{string}}V_j^{\text{bat}}, \quad j = 1,\,2,\,\ldots,\,N,\label{eqn:opt_layer2_contraint3} \\
& p^{(2)}_k \le p^{(2)}_{\text{max}}, \quad k = 1,\,2,\,\ldots,\,N-1,\label{eqn:opt_layer2_contraint4}\\
& p^{(1)}_{i} \le {p^{(1)}_{i}}^{*}, \quad i = 1,\ldots,M \label{eqn:opt_layer2_contraint5},
\end{align}
where $K^{(2)}_j$ is the index set of the Layer 2 converters whose inputs are connected to the $j^{\text{th}}$ battery. In this optimization problem, the decision variables are the power processed by the $N-1$ Layer 2 converters, represented by $p^{(2)}_1 \ldots p^{(2)}_{N-1}$.  The optimization objective (\ref{eqn:opt_layer2_obj}) maximizes the total power transferred to the bus. Constraint (\ref{eqn:opt_layer2_contraint1}) are the battery input and output power limits. $\delta P_j$ represents the power uncertainty of the $j^{\text{th}}$ battery. Constraint (\ref{eqn:opt_layer2_contraint2}) is the power conservation law for each battery. Constraint (\ref{eqn:opt_layer2_contraint3}) expresses the direct power transfer from the $j^{\text{th}}$ battery to the bus. Constraint (\ref{eqn:opt_layer2_contraint4}) indicates that the power ratings for all Layer 2 converters are identically equal to $p_{\text{max}}^{(2)}$. Constraint (\ref{eqn:opt_layer2_contraint5}) suggests that the Layer 1 converter ratings are kept fixed during the layer 2 power processing design. 

% The average battery power utilization from the optimal set is designated to that converter rating. Hence, a particular Layer 2 converter rating will have this battery power utilization metric.

% \begin{figure} % To do:  Block diagram of Layer 2 design
%     \centering
%     \includegraphics{}
%     \caption{Caption}
%     \label{fig:my_label}
% \end{figure}
%\Figure[t!](topskip=0pt, botskip=0pt, midskip=0pt)[width = 16cm]{"layer2.pdf"}{Power flow design for Layer 2 power processing is subsequent to the  design of the Layer 1 converter ratings and interconnection.  In the linear programming formulation for power capability, the ratings for the Layer 2 converters are constraints.  From Monte Carlo trials, the Expected Battery Utilization for the 2-BESS can be obtained.\label{fig:layer2}}

\Figure[t!](topskip=0pt, botskip=0pt, midskip=0pt)[width=\linewidth]{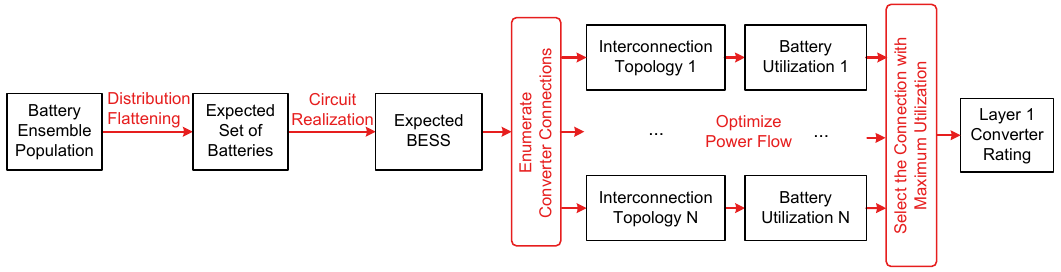}{\label{fig:layer1_production}The design flow of the power ratings of Layer 1 converters. The expected set of batteries is derived from the Distribution Flattening of the battery supply set. We enumerate the interconnections of expected batteries set with Layer 1 converters. For every interconnection, the optimal power flow is found using linear programming. At the maximum utilization, the converters will have optimal interconnection and power flow.}

\Figure[t!](topskip=0pt, botskip=0pt, midskip=0pt)[width=\linewidth]{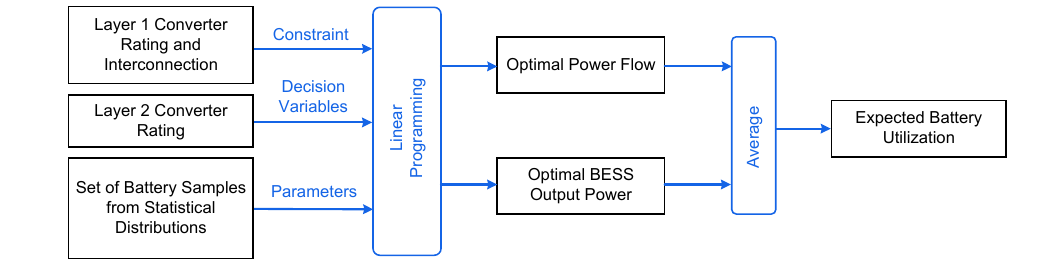}{Relationship between expected battery utilization and Layer 2 power processing.  Layer 2 power processing is chosen after the design of the Layer 1 converter ratings and interconnection.  In the linear programming formulation for power capability, the ratings for the Layer 2 converters are constraints. From Monte Carlo trials, the expected battery utilization for the 2-BESS can be obtained.\label{fig:layer2}}

%\Figure[t!](topskip=0pt, botskip=0pt, midskip=0pt)[width = 16cm]{"actualization.png"}{Production of a 2-BESS unit.\label{fig:actualization}}

\Figure[t!](topskip=0pt, botskip=0pt, midskip=0pt)[width=\linewidth]{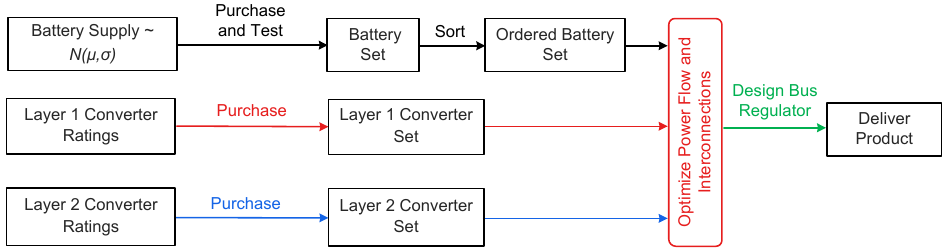}{Production flow of a 2-BESS unit: (i) Layer 1 and 2 converter ratings are predetermined and purchased in volume; (ii) The batteries are tested and ordered in the series string; (iii) Layer 1 interconnections are optimized; (iv) Layer 2 is connected; and (v) Power flow is optimized. 2-BESS units are optimized for heterogeneity using power converters that are purchased at economies of scale.\label{fig:actualization}}

\subsection{Allocation in and Production \\ of 2-BESS Units}
\begin{comment}
A set of batteries are drawn from a supply. The batteries are sorted from the lowest to highest power capability and connected in series.  Power converters contribute a significant portion of the cost of a 2-BESS unit.  Economies of scale can alleviate this cost and is a directive and, from the perspective of optimization, can be considered an objective.  Economies of scale can be accomplished by using fewer types of converters, i.e. using as many converters with the same rating as possible.
This is a goal in Section~\ref{sec:converter-sets}.  
\end{comment}

In allocation, both Layer 1 and Layer 2 converters are purchased production quantities, which are determined by the prediction of the 2-BESS product demand. Both Layer 1 and Layer 2 power converters are connected according to Section \ref{sec:layer1design} and \ref{sec:layer2design}, respectively.

The production flow of a 2-BESS unit is illustrated in Fig.\,\ref{fig:actualization}. A battery set for a particular 2-BESS unit is acquired from the battery supply.  For a power-intensive application, the batteries are then tested and evaluated for power capability. The assembly of the 2-BESS unit proceeds by sorting and then connecting the batteries so they are ordered from the lowest to highest capability.  Using the battery capability information from testing, the interconnection of Layer 1 converters (pre-purchased with a specified rating) is determined using the method outlined in Section\,\ref{sec:layer1design}. Both Layer 1 and 2 converters are then connected to the batteries and power is optimized for the application.  Then, the 2-BESS unit is tested and validated for performance and safety before being delivered to the customer.

% To do:
%Deployment design  -- flow diagram

%Flow diagram

%Supply Delivery/Selection

%Measure capabilities [ref]

%Sort

% Section:  Analysis of Alternative Strategies: Trying to overcome heterogeneity through abundance of supply.

The optimal power flows are recalculated using linear programming during operation with the actual battery power capabilities in the 2-BESS unit and the output load as constraints. As we show in Section \ref{sec:results}, the design method for LS\nobreakdash-HiPPP results in a better cost-performance tradeoff as it relates to battery power utilization and aggregate converter rating compared to the state-of-the-art approaches. The LS\nobreakdash-HiPPP architecture performs well with a sparse set of Layer 1 converters at moderate power ratings and a dense set of Layer 2 converters at lower power ratings.

\section{Results and Discussion}\label{sec:results}
LS\nobreakdash-HiPPP enables tradeoffs in performance and price not previously possible using conventional methods.  The LS\nobreakdash-HiPPP design methods have tractable complexity in the optimization on the order of $10^4$ linear programming iterations with 200 variables for nine batteries and three Layer 1 converters, which can be performed on a small computing cluster.  The specific LS\nobreakdash-HiPPP realization is illustrated in Fig.\,\ref{fig:LS-HiPPP2}, which consists of nine heterogeneous batteries connected in series with three Layer 1 power converters and nine Layer 2 converters.

The battery power capability was modeled as a Gaussian distribution with a normalized expected power $\hat{\mu}_p = 1$ with the {\em power heterogeneity} represented by the standard deviation $\hat{\sigma}_p$ normalized to the expected power. 
The aggregate converter rating $\mathcal{\hat{R}}_p$ is normalized to the average aggregate intrinsic battery power $\bar{P_I} $,
\begin{align}
    \mathcal{\hat{R}}_p & = \frac{\sum\limits_{i = 1}^{M} {p^{(1)}_{i}}^{*}+ \sum\limits_{i = 1}^{N-1} p^{(2)}_{\text{max}}}{\bar{P_I}},\\
    \bar{P_I} & \triangleq \sum\limits_{i = 1 }^{N}\bar{P_i}.
\end{align}
With a constant \$/kW presupposition, aggregate converter rating is a proxy for converter cost.

\subsubsection{Partial Power Processing vs. Full Power Processing} \label{sec:ppp_vs_fpp}
A tradeoff between the expected battery power utilization and the normalized aggregate converter rating was investigated using Monte Carlo simulations.  We compare LS\nobreakdash-HiPPP with two state-of-the-art approaches for heterogeneous batteries in a 2-BESS: FPP and C-PPP.

\Figure[t!](topskip=0pt, botskip=0pt, midskip=0pt)[width=8.5cm]{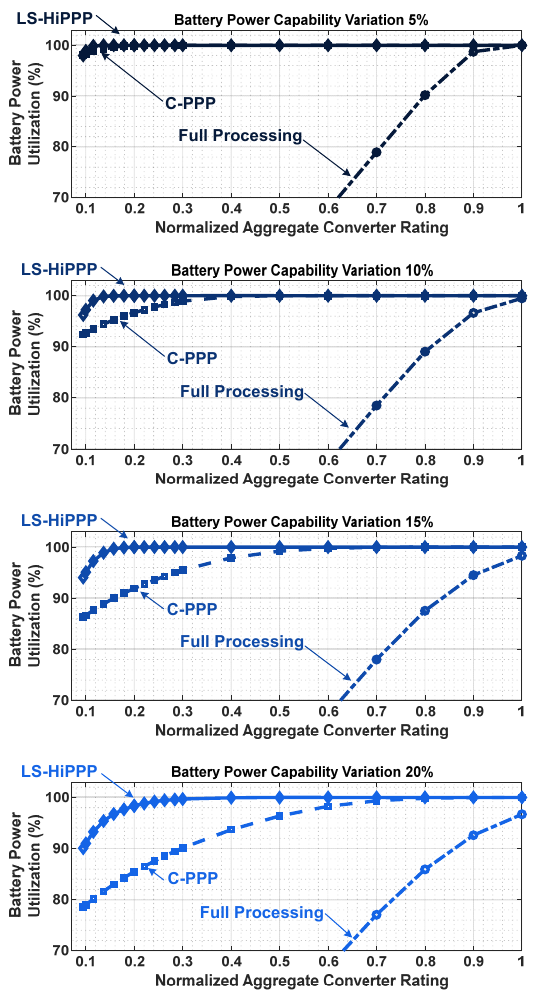}{
Comparison of battery utilization as a function of aggregate converter rating $\mathcal{\hat{R}}_p$ for full power processing vs. partial power processing: LS-HiPPP and C-PPP.  Partial power processing architectures perform significantly better than FPP. \label{fig:battu-v-rating-v-fpp}}

We can observe from Fig.\,\ref{fig:battu-v-rating-v-fpp} that partial power processing architectures perform significantly better even at higher heterogeneity (20\%) and low converter ratings (20\%).  The battery utilization for full processing is very nearly linear to the converter rating; this is the case because all of the battery power must be processed by the power converter, hence making the power converter rating the limiting factor.

Fig.\,\ref{fig:battu-v-rating-v-fpp} shows the results for heterogeneity in the battery power capability with a standard deviation of 20\% of the mean.  Because the maximum output power is equal to the aggregate power converter rating for FPP, the utilization curve increases linearly with converter rating, resulting in the most costly option for converter cost per unit power capability for 2-BESS.  One-layer or C-PPP shows a better tradeoff than FPP because only the mismatch power is processed.

\subsubsection{LS\nobreakdash-HiPPP vs. C-PPP}
LS\nobreakdash-HiPPP shows the best tradeoff for battery power utilization and converter cost.  This particular LS\nobreakdash-HiPPP design, shown in Fig.\,\ref{fig:LS-HiPPP2}, uses only three Layer 1 power converters with interconnections and ratings that are optimal for the structure of the statistical distribution of the battery supply.  The Layer 2 converters are low power and are designed to accommodate the deviations from the statistical distribution.  As a point of comparison in Fig.\,\ref{fig:battu-conv-2}, LS\nobreakdash-HiPPP has an expected battery power utilization of 95\% with only 15\% of the output power processed as opposed to 81\% for conventional PPP and 15\% for full processing.  For the same utilization, LS\nobreakdash-HiPPP requires approximately one-fifth of the power converter rating of FPP with the same corresponding reduction in power converter cost, assuming constant \$/kW.

\Figure[t!](topskip=0pt, botskip=0pt, midskip=0pt)[width = 8.5cm ]{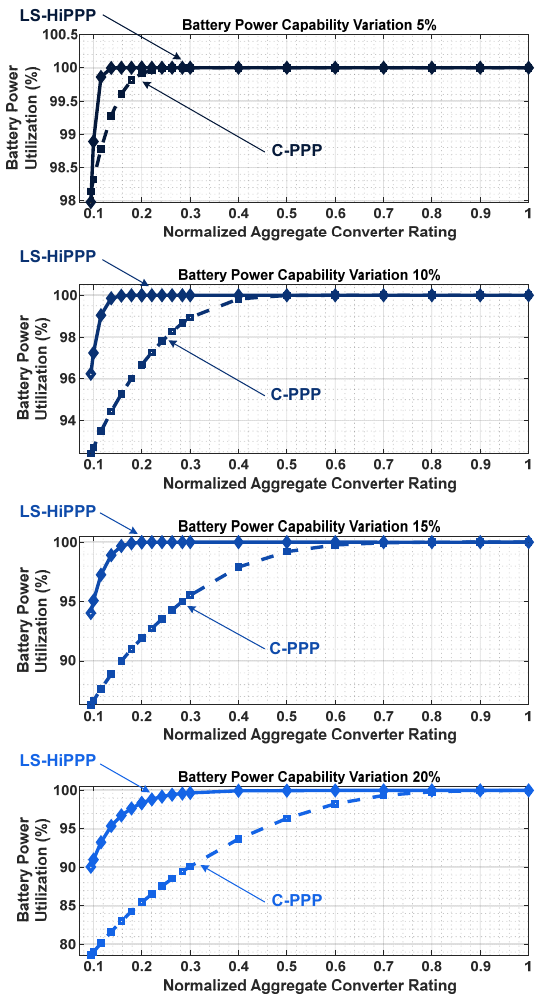}{Comparison of battery utilization as a function of aggregate converter rating for two partial power processing architectures: \mbox{LS-HiPPP} and C-PPP.  \mbox{LS-HiPPP} performs significantly better than C-PPP, needing much lower converter ratings for the same performance.  At \mbox{$\mathcal{\hat{R}}_p = 15\%$} converter rating and 20\% heterogeneity, LS-HiPPP has an battery power utilization of over 95\% as opposed to 81\% for C-PPP.\label{fig:battu-conv-2}}

\subsubsection{Effect of Heterogeneity on Battery Utilization}
Heterogeneity decreases battery utilization in all second-use BESSs. This is because power converter ratings are selected for a particular design point and pre-allocated with a directive towards economies of scale. We have observed from Section\,\ref{sec:ppp_vs_fpp} that partial power processing architectures offer the best tradeoff between battery utilization and converter cost.

In this section, we compare the battery utilization for LS\nobreakdash-HiPPP with C-PPP as battery heterogeneity increases.

\begin{figure}[ht]
\centering
\begin{subfigure}[Comparison of battery utilization as a function of battery heterogeneity between LS\nobreakdash-HiPPP and C-PPP for $\mathcal{\hat{R}}_p = 20\%$. LS\nobreakdash-HiPPP performs better at all levels of heterogeneity.] 
    {\centering
    \includegraphics[width = 8cm]{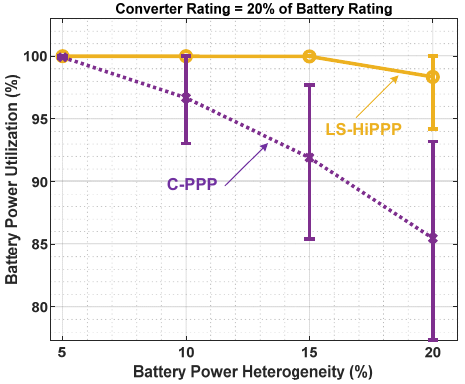}
    \label{fig:u_heter_b}
    }
\end{subfigure}
\newline
\begin{subfigure}[Comparison of standard deviation of the battery power utilization as a function of battery heterogeneity between LS\nobreakdash-HiPPP and C-PPP for $\mathcal{\hat{R}}_p = 20\%$. LS\nobreakdash-HiPPP performs better at all levels of heterogeneity.]
    {
    \centering
    \includegraphics[width = 8cm]{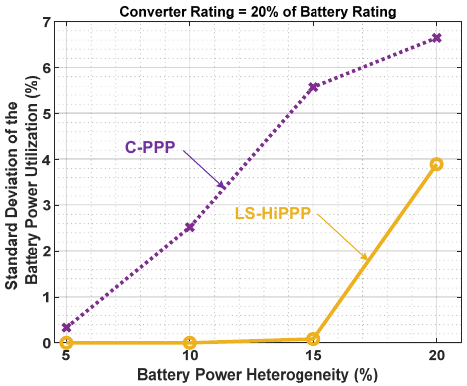}
    \label{fig:std_heter_b}
    }
\end{subfigure}
\caption{Comparison of battery utilization between LS\nobreakdash-HiPPP and C-PPP for $\mathcal{\hat{R}}_p = 20\%$.}
\end{figure}

As illustrated in Fig.\,\ref{fig:u_heter_b} and Fig.\,\ref{fig:std_heter_b}, for \mbox{$\mathcal{\hat{R}}_p = 20\%$}, LS\nobreakdash-HiPPP always performs better than C-PPP in the expected value and standard deviation of battery power utilization. Lower standard deviation in the 2-BESS battery power utilization means that the sensitivity to individual battery variation is less.  
The performance of C-PPP falls more drastically at higher battery heterogeneity.  In other words, not only is the expected battery power utilization, but also the variation from BESS unit to BESS unit across production is significantly lower for LS-HiPPP.

% \Figure[t!](topskip=0pt, botskip=0pt, midskip=0pt)[width = 8cm]{u_heter.png}{Comparison of battery utilization as a function of battery heterogeneity between LS\nobreakdash-HiPPP and C-PPP for $\mathcal{\hat{R}}_p = 20\%$. LS\nobreakdash-HiPPP performs better \label{fig:battu-heter} at all levels of heterogeneity.}

% \begin{figure}
%     \centering
%     \includegraphics[width = 8cm]{u_heter_b.png}
%     \caption{Battery power heterogeneity to the standard deviation of the battery power utilization \red{need descriptions in texts}}
%     \label{fig:u_heter_b}
% \end{figure}

\subsection{System Efficiency}\label{sec:sys_eff}
The system efficiency is the ratio of the power delivered by the output of the 2-BESS to the sum of the power delivered by each individual battery within the 2-BESS, \footnote{The system efficiency refers to the electrical system efficiency and does not include the auxiliary power consumption} \cite{Rancilio2020}, such as cooling and control.
\begin{align}
    \eta_{\text{sys}} &= \frac{P_{\text{out}} - P_{\text{loss}}}{P_{\text{out}}}\nonumber \\ 
         &= 1 - \frac{P_{\text{loss}}}{P_{\text{out}}}\nonumber\\
         &= 1 - \frac{P_{\text{proc}}}{{P_{\text{out}}}} (1- \eta_{\text{conv}}), \label{eqn:sys-eff}
\end{align}
where $P_{\text{loss}}$ represents the converter loss, $P_{\text{proc}}$ represents the processed power by the converters, and $\eta_{\text{conv}}$ is the efficiency of the converters.  High system efficiency means lower overall power losses which also means a lower cost of cooling and thermal management.  However, high efficiency converters have higher cost.  If high system efficiency can be achieved by processing less power or compromising output power capability, then lower cost converters can be used without increasing the cost of cooling. It can be observed from Fig.\,\ref{fig:processpower_eff_syseff} that as processed power is reduced, system efficiency increases and is less sensitive to converter efficiency.

Compared to FPP and C-PPP, which are the state-of-the-art architectures, LS-HiPPP processes the least amount of power. From Fig.\,\ref{fig:battu-v-rating-v-fpp}, for the same power output (e.g. 95\% of the battery power), LS-HiPPP processes only 14\% of the battery power. C-PPP processes 46\% and FPP processes 100\% of the battery power. Based on (\ref{eqn:sys-eff}) and Fig.\,\ref{fig:processpower_eff_syseff}, the electrical system efficiency is 98\% for LS-HiPPP, 93\% for C-PPP, and 85\% for FPP..\!\footnote{A typical efficiency of low-cost converters is 85\%.} Therefore, LS-HiPPP shows the best efficiency compared to C-PPP and FPP.

% in terms of overall power/sum, battery power and converter losses, battery power and converter efficiency...
\Figure[t!](topskip=0pt, botskip=0pt, midskip=0pt)[width = 8cm ]{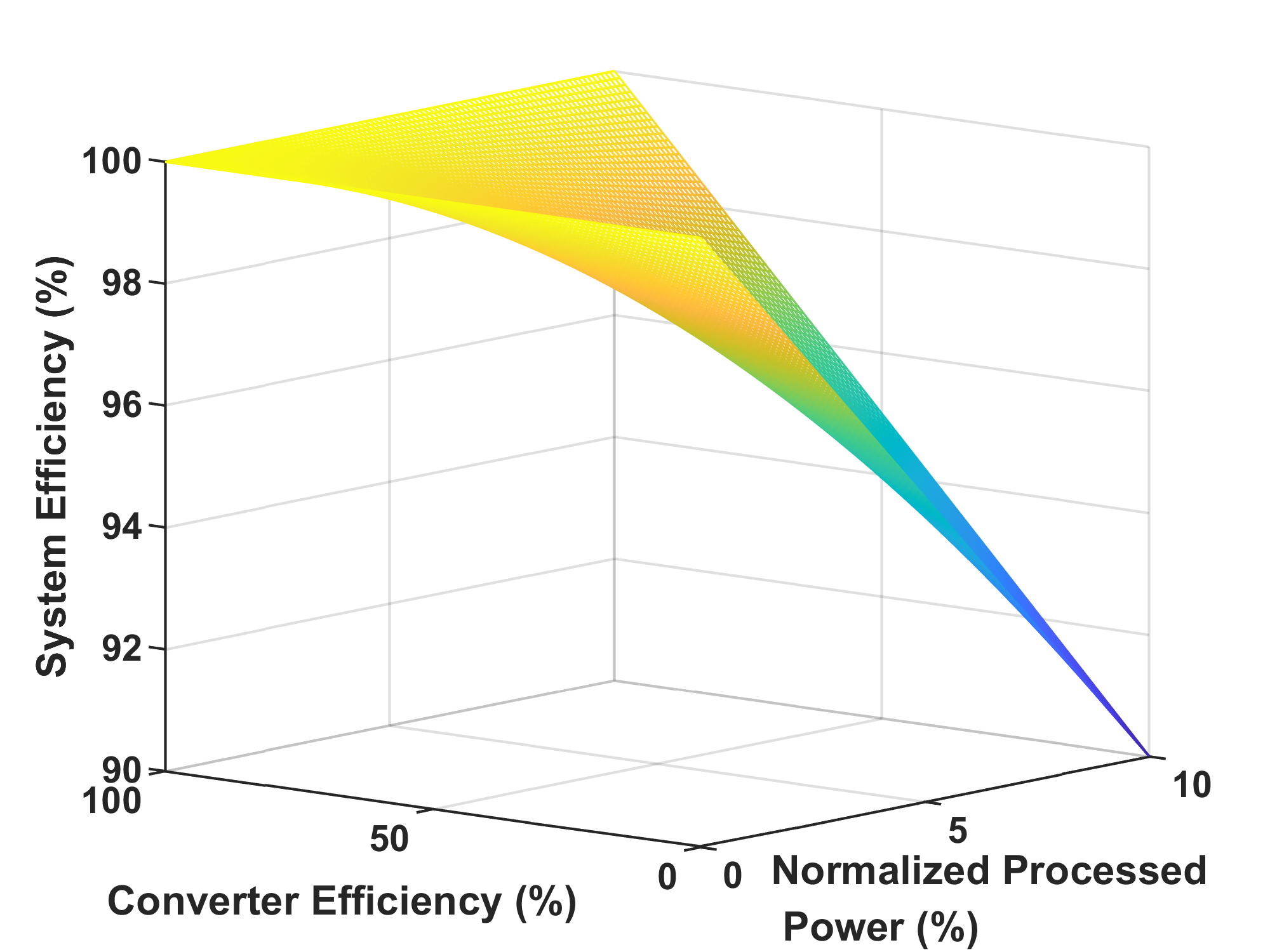}{3-d plot of system efficiency vs. normalized processed power vs. power converter efficiency. Smaller processed power can result in high system efficiency despite lower power converter efficiency (lower cost power converters). \label{fig:processpower_eff_syseff}}

\subsubsection{LS\nobreakdash-HiPPP vs. C-PPP}
In comparing the system efficiency when using low-cost converters with a meager efficiency of 85\%, Fig.\,\ref{fig:eff-conv-2} shows that at \mbox{$\mathcal{\hat{R}}_p = 15\%$} converter power rating, LS\nobreakdash-HiPPP has a system efficiency of 98.8\%, while C-PPP has 97.8\% and FPP only 85\% (not shown in the figure) because 100\% of the power is processed by the converters.

\Figure[t!](topskip=0pt, botskip=0pt, midskip=0pt)[width = 8.5cm ]{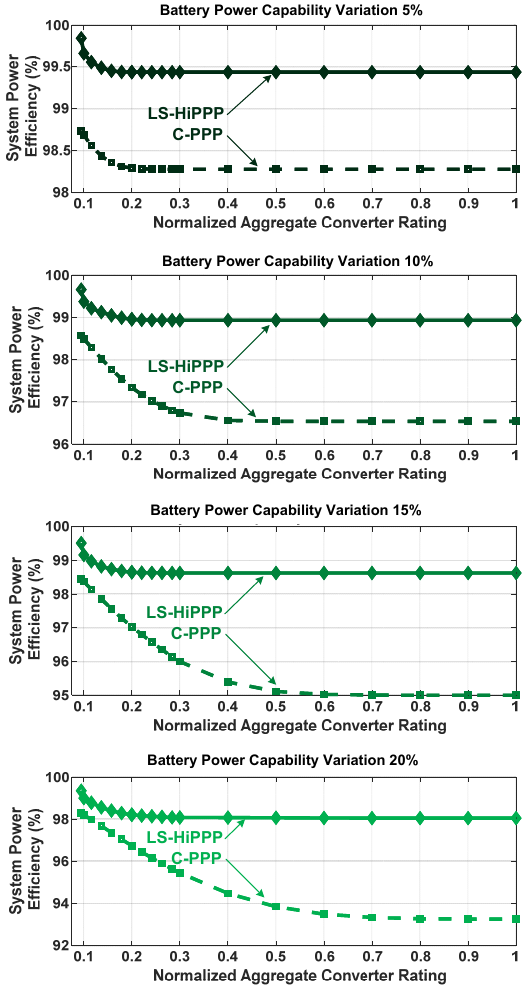}{Comparison of system efficiency as a function of aggregate converter power rating \mbox{$\mathcal{\hat{R}}_p$} for two partial power processing architectures: \mbox{LS-HiPPP} and C-PPP. \mbox{LS-HiPPP} has a higher system efficiency for all cases of heterogeneity and converter rating $\mathcal{\hat{R}}_p$.  System efficiency is especially impactful in reducing thermal management and cooling costs.\label{fig:eff-conv-2}}

System efficiency has a significant impact on the cost of thermal management.  At high efficiencies even a single digit improvement in efficiency is impactful, e.g. at 99\%, a 1\% decrease in efficiency doubles the requirement for cooling.  For example, in a 500 kW BESS operating at 99\% efficiency, a 1\% decrease in efficiency corresponds to an increase in required heat removal from 5\,kW to 10\,kW.

\subsubsection{Effect of Heterogeneity on System Efficiency}
Battery heterogeneity decreases system efficiency in partial power processing systems.  Partial power processing systems are designed to mainly process the mismatch power; heterogeneity increases this mismatch power that the power converters need to process.  Fig.\,\ref{fig:eff-conv-2} shows that as battery heterogeneity becomes higher, the system power efficiency decreases. From Figs.\,\ref{fig:eff-conv-2} and \,\ref{fig:eff-heter}, LS\nobreakdash-HiPPP has higher system efficiency than C-PPP in all cases. Fig.\,\ref{fig:eff-heter} illustrates the specific case for \mbox{$\mathcal{\hat{R}}_p = 20\%$}; as battery heterogeneity is greater, the system efficiency decreases.

\Figure[t!](topskip=0pt, botskip=0pt, midskip=0pt)[width = 8cm ]{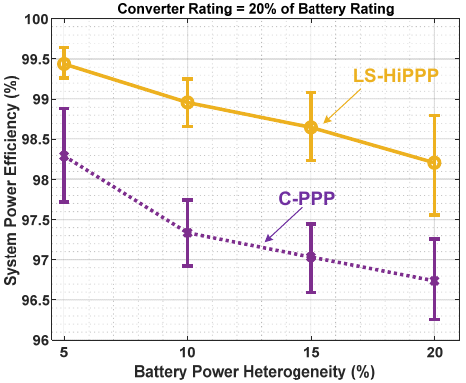}{Comparison of system power efficiency as a function of battery heterogeneity between \mbox{LS-HiPPP} and \mbox{C-PPP} for \mbox{$\mathcal{\hat{R}}_p = 20\%$}.
\mbox{LS-HiPPP} has a higher system efficiency for all cases of heterogeneity.  System efficiency is especially impactful in reducing thermal management and cooling costs.\label{fig:eff-heter}}

\subsection{Benefits of Lower Processed Power for LS\nobreakdash-HiPPP}
Processed power is the aggregate power flow through the power converters and is normalized by the aggregate intrinsic battery power $\bar{P}_I$
\begin{align}
   \hat{P}_{\text{proc}} =  \frac{P_{\text{proc}}}{\bar{P}_I}.
\end{align}
The normalized output power is
\begin{align}
    \hat{P}_{\text{out}} = \frac{P_{\text{out}}}{\bar{P}_I},
\end{align}
where $P_{\text{out}}$ is the output power and $\bar{P}_I$ is the aggregate intrinsic battery power.

\Figure[t!](topskip=0pt, botskip=0pt, midskip=0pt)[width = 8.5cm ]{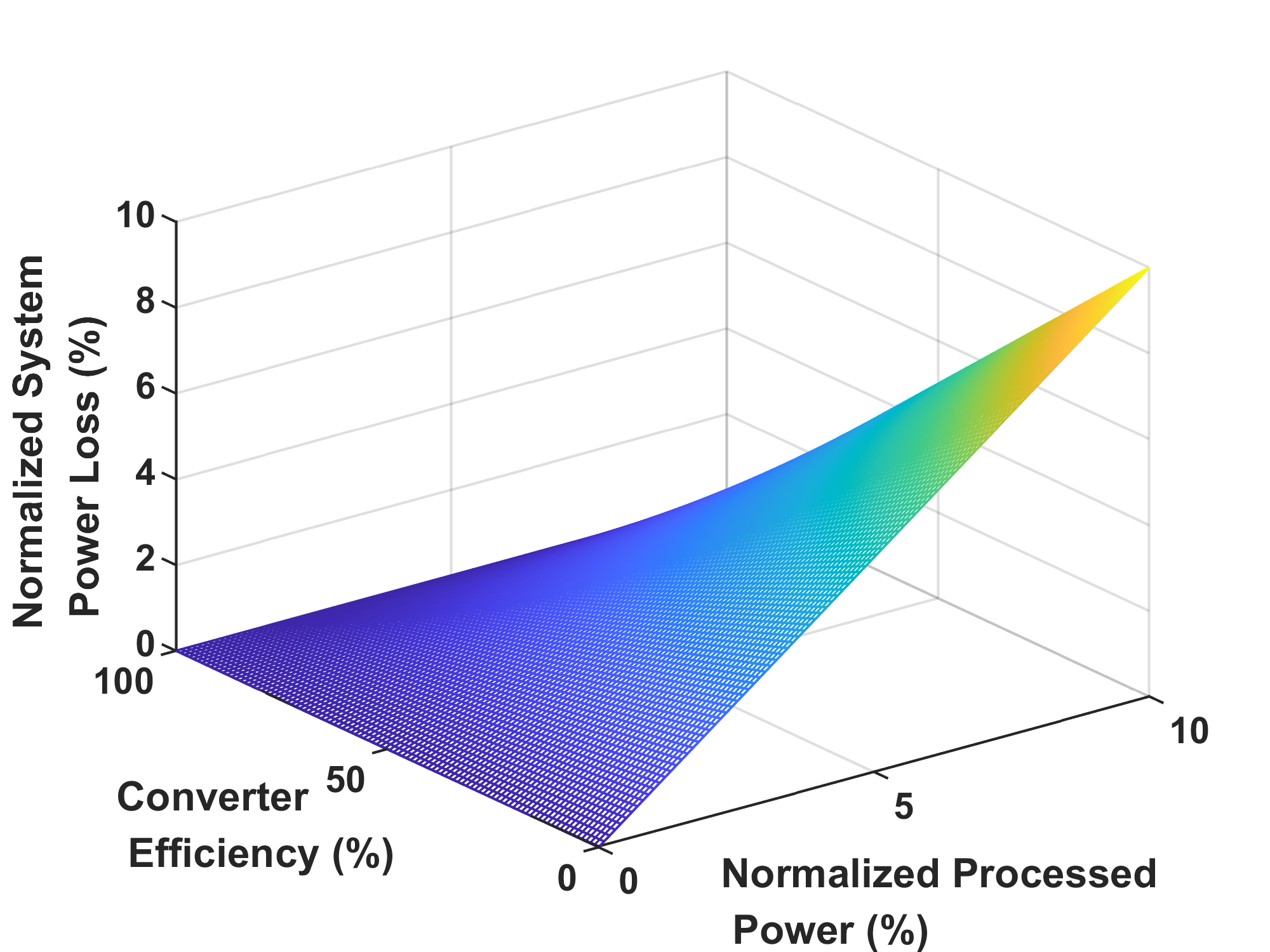}{
3-d plot of normalized power loss vs. normalized processed power vs. power converter efficiency. Lower processed power results in lower power loss despite lower efficiency (lower cost) power converters.  Lower power loss means lower cost of thermal management and cooling.\label{fig:processpower_eff_powerloss} }

\Figure[t!](topskip=0pt, botskip=0pt, midskip=0pt)[width = 8cm ]{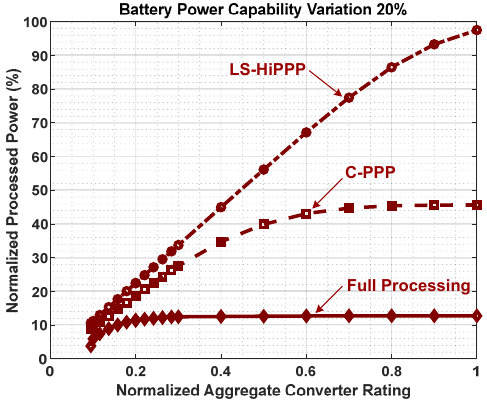}{Converter rating \mbox{$\mathcal{\hat{R}}_p$} vs. processed power at 20\% heterogeneity. \mbox{LS-HiPPP} has lower processed power because the Layer 1 sparse converters are optimally placed and allow for a better overall power flow optimization. \label{fig:processedpower_conv}}  

There are several consequences to having an architecture with low processed power, as illustrated in Fig.\,\ref{fig:processpower_eff_powerloss}. The first is lower requirements and subsequently lower cost for cooling.  Second, lower processed power means that converters with lower ratings and hence lower costs are required.  In comparing the two partial power processing architectures, LS\nobreakdash-HiPPP and C-PPP, with FPP in Fig.\,\ref{fig:processedpower_conv}, LS\nobreakdash-HiPPP offers the lowest processed power for any choice of converter rating.

\Figure[t!](topskip=0pt, botskip=0pt, midskip=0pt)[width = 8cm ]{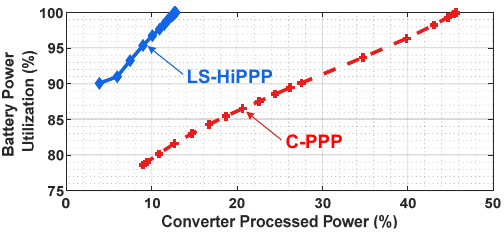}{Tradeoff frontier of battery power utilization and converter processed power. Even at 100\% battery utilization, the processed power for \mbox{LS-HiPPP} does not exceed 13\%; the caveat is that converter rating \mbox{$\mathcal{\hat{R}}_p$} increases with battery utilization although not as steeply as FPP and \mbox{C-PPP}. \label{fig:frontier-pconv}}

In comparing partial power processing architectures in Fig.\,\ref{fig:frontier-pconv}, LS\nobreakdash-HiPPP delivers significantly high battery utilization at low processed power.  Not only does LS\nobreakdash-HiPPP have higher battery utilization, but it can also do so at high system efficiency as illustrated in Fig.\,\ref{fig:frontier-peff}.

\Figure[t!](topskip=0pt, botskip=0pt, midskip=0pt)[width = 8cm ]{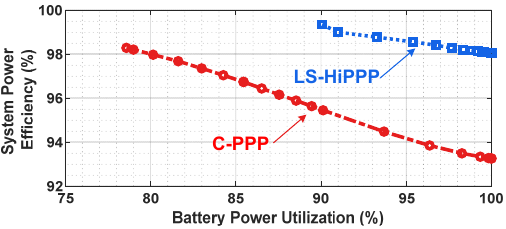}{Tradeoff frontier of battery power utilization and system efficiency. At high battery power utilization, \mbox{LS-HiPPP} maintains high system power efficiency despite using 85\% efficient power converters.  Comparatively, C-PPP has a higher penalty in system power efficiency at high battery power utilization. System efficiency is especially impactful in reducing thermal management and cooling costs. \label{fig:frontier-peff}}

\subsection{Power Derating} \label{sec:power_derating}
The {\it power derating} in the context of this paper is the power capability of a particular 2-BESS within a 99.85\,\% confidence level when the batteries are drawn from a statistically distributed supply.
Given a 2-BESS with a power capability distribution described by expected value $\mu_{2b}$ and standard deviation $\sigma_{2b}$, the $3\sigma$ power derating can be calculated by
\begin{align}
    \mathcal{D}_f = \frac{\mu_{2b} - 3\sigma_{2b}}{\mu_{2b}}.
\end{align}

\begin{figure}
    \centering
    \includegraphics[width = 8cm]{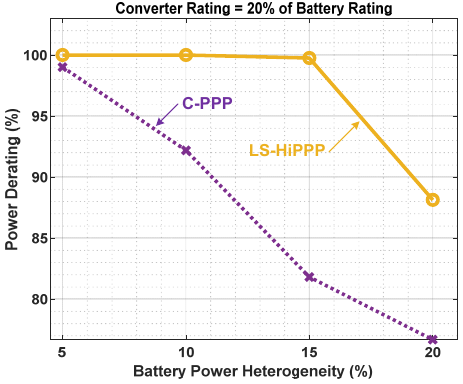}
    \caption{Comparison of $3\sigma$ power derating as the function of battery power heterogeneity between LS-HiPPP and C-PPP at 20\% normalized aggregate power converter rating.}
    \label{fig:power_derating}
\end{figure}

We compare the power derating for LS-HiPPP with C-PPP as battery heterogeneity increases. As illustrated in Fig.\,\ref{fig:power_derating}, for $\mathcal{\hat{R}}_p = 20\%$, LS-HiPPP outperforms C-PPP. The power derating of C-PPP decreases more significantly at greater battery heterogeneity.

As $\mathcal{\hat{R}}_p$ becomes higher, the power derating increases. LS\nobreakdash-HiPPP has higher power derating than C-PPP in all cases.
The power derating of LS\nobreakdash-HiPPP increases more drastically with $\mathcal{\hat{R}}_p$ than C\nobreakdash-PPP.  In other words, to confidently obtain 100 kW from a production 2-BESS, one must purchase a 2-BESS rated at 129.87\,kW for C-PPP, but a lower rated 113.64\,kW 2-BESS at a lower cost for LS-HiPPP.

\begin{figure}
    \centering
    \includegraphics[width = 8cm]{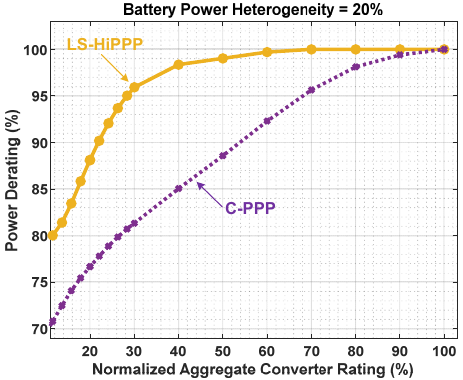}
    \caption{Comparison of $3\sigma$ power derating as the function of normalized aggregate power converter rating between LS-HiPPP and C-PPP at 20\% battery power heterogeneity.}
    \label{fig:power_derating_reverse}
\end{figure}

\subsection{Power Captured Value}
The expected financial value of a 2-BESS capability is determined by both derating factor $\mathcal{D}_f$ and power utilization $\mathcal{U}_P$. The power captured value is defined as 
\begin{equation}
    \mathcal{C}_v \triangleq \mathcal{D}_f\,\mathcal{U}_P,
\end{equation}
where $\mathcal{C}_v$ is in the unit of kW per kW-intrinsic.

\begin{figure}
    \centering
    \includegraphics[width = 8cm]{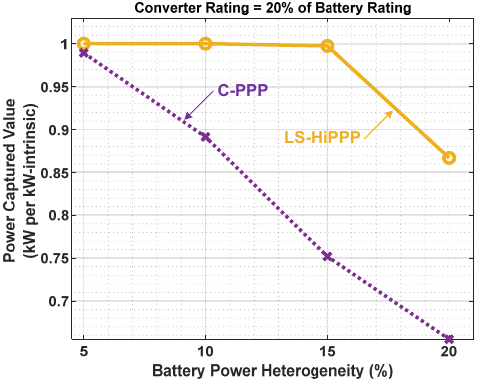}
    \caption{Comparison of captured value as the function of battery power heterogeneity between LS-HiPPP and C-PPP at 20\% normalized aggregate power converter rating.}
    \label{fig:captured_value}
\end{figure}

We  compare  the  power captured value for LS-HiPPP  with  C-PPP  as battery  heterogeneity increases. For $\mathcal{\hat{R}}_p = 20\%$, LS-HiPPP outperforms C-PPP, as illustrated in Fig.\,\ref{fig:captured_value}. 
The performance of C-PPP decreases more significantly at greater battery heterogeneity.

\begin{figure}
    \centering
    \includegraphics[width = 8cm]{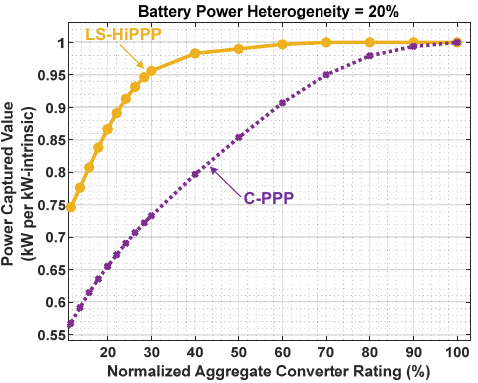}
    \caption{Comparison of captured value as the function of normalized aggregate power converter rating between LS-HiPPP and C-PPP at 20\% battery power heterogeneity.}
    \label{fig:captured_value_reverse}
\end{figure}

Fig.\,\ref{fig:captured_value_reverse} is the true tradeoff curve between the risk value pricing of second-use batteries and the cost of power converters. 
As $\mathcal{\hat{R}}_p$ becomes higher, the power captured value increases. LS\nobreakdash-HiPPP has higher power captured value than C-PPP in all cases. The power captured value of LS\nobreakdash-HiPPP increases more drastically with $\mathcal{\hat{R}}_p$ than C\nobreakdash-PPP.


\subsection{Discussion}
% From numerical data in Fig.}\,\ref{fig:battu-conv-2}, given 20\% battery power capability variation, for the same power output (e.g., 95\% of the battery capability), LS-HiPPP processes only 14\% of the power output. C-PPP processes 46\% and FPP processes 100\% of the power output. Based on the equation} (\ref{eqn:sys-eff}) and Fig.\,\ref{fig:processpower_eff_syseff}, the electrical system efficiency for power converters with 85\% efficiency \footnote{typical efficiency for low-cost converters} is 98\% for LS-HiPPP, 93\% for C-PPP and 85\% for FPP.
% This shows the LS-HiPPP has the best efficiency compared to C-PPP and FPP.}
\subsubsection{Microgrid Applications}
2-BESS is useful for stabilizing the voltages of microgrids. Compared to other power processing architectures, LS-HiPPP is lower in cost and higher in power capability. The LS-HiPPP can provide energy to the grid quickly when the voltage droops. The bus voltage regulator can quickly match the output of 2-BESS with that required for the grid voltage.

2-BESS using LS-HiPPP can operate as a grid-forming unit and support an island microgrid \cite{Shahparasti2022} when the bus voltage regulator is controlled to make the BESS output behave as a stiff voltage source together with an inverter.

\subsubsection{Sensor Inaccuracy}
A practical challenge of the proposed method is sensor inaccuracy.
Measurement and random errors typically follow a Gaussian distribution.
Typically, the current of each battery is usually measured to within 0.1\% accuracy. The voltage of each battery is usually measured to within 0.025\% accuracy.
The current and voltage measurement accuracy results in the power error measured at the terminal of each battery to be approximately 0.1\%.
The power and energy metering accuracy for billing is within 2\% \cite{Kong2022}.
We use this utility metering accuracy standard to compare with the 2-BESS accuracy.
Given a 2-BESS with $N$ batteries, the power capability percentage error of the 2-BESS $e_{_\text{2Bess}}$ can be bounded over the range of individual battery power capabilities as
\begin{align}
    0.1\% \times \frac{1}{\sqrt{N}} \le e_{_\text{2Bess}} \le 0.1\%.
\end{align}
This error $e_{_\text{2Bess}}$ is better than the utility metering accuracy.

\subsubsection{Practical Challenges}
A potential challenge of the proposed method is that plenty of data is needed to acquire adequate statistics. Machine learning and artificial intelligence (AI) can assist in solving the practical challenges \cite{Liu2022, Liu2022a}.

\subsubsection{Cost Analysis}
Power converters and second-use batteries dominate the cost of 2-BESS\cite{Neubauer2015}. We denote the per kW cost of power converters by a constant $c_{_\text{conv}}$ and the per kWh  cost of second-use batteries by a constant $c_{_\text{batt}}$.
\paragraph{Cost Analysis Based on Expected Utilization}
We first show a simple analysis based on expected utilization. In this case, to build a 2-BESS with 100\,kW (at 1\,C Rate) power output capability, the cost of 2-BESS $c_{_\text{2Bess}}$ is
\begin{align} \label{eqn:total_cost}
    c_{_\text{2Bess}} = \frac{100 \,\text{kW}}{\mathcal{U}_P(\mathcal{\hat{R}}_p)}\; c_{_\text{batt}} +
    \frac{100\,\text{kWh}}{\mathcal{U}_P(\mathcal{\hat{R}}_p)}  \; \mathcal{\hat{R}}_p c_{_\text{conv}},
\end{align}
where $\mathcal{U}_P$ is the power utilization of the batteries.
The cost of 2-BESS $c_{\text{2Bess}}$ varies with the power converter rating $\mathcal{\hat{R}}_p$. The minimal cost is obtained if the marginal cost of power converters equals the negative of the marginal cost of second\nobreakdash-use batteries
\begin{align} \label{eqn:opt_cost}
    c_{_\text{batt}} \, \frac{\partial \mathcal{U}_P}{\partial \mathcal{\hat{R}}_p} \Bigg\rvert_{\mathcal{\hat{R}}_p^*} = c_{_\text{conv}} \, \left( \mathcal{U^{*}_P}- \frac{\partial \mathcal{U}_P}{\partial \mathcal{\hat{R}}_p} \Bigg\rvert_{\mathcal{\hat{R}}^{*}_p} \mathcal{\hat{R}}^{*}_p\right), 
\end{align}
where $\mathcal{U^{*}_P}$ is the optimal power utilization of the batteries and $\mathcal{\hat{R}}^{*}_p$ is optimal power converter rating.
\begin{table}[tbp]
    \caption{Cost Comparison of Three Power Processing Architectures Using Expected Utilization}
    \label{table:cost_pp_compare}
    \centering
    \begin{tabular}{ccc}
    \toprule
        \textbf{} &\textbf{Batteries (\$)} &\textbf{Converters (\$)} \\
        \midrule
        \textbf{LS-HiPPP} & 105.0$\,c_{\text{h}}$ & 14.2$\,c_{\text{h}}$  \\
        \midrule
        \textbf{C-PPP} & 124.3$\,c_{\text{h}}$ & 14.9$\,c_{\text{h}}$  \\
        \midrule
        \textbf{FPP} & 100 $\,c_{\text{h}}$ & 100 $\,c_{\text{h}}$ \\
    \bottomrule
    \end{tabular}
\end{table}
For example, if the cost of batteries $c_{_\text{batt}}$ and power converters $c_{_\text{conv}}$ are both equal to $c_{\text{h}}$,
by using Fig.\,\ref{fig:battu-v-rating-v-fpp} and solving (\ref{eqn:total_cost}) and (\ref{eqn:opt_cost}), we can compare the optimal cost of different power processing architectures as shown in Table\,\ref{table:cost_pp_compare}.

The optimal power converter rating $\mathcal{\hat{R}}^{*}_p$ for C-PPP is 14.9\% while $\mathcal{\hat{R}}^{*}_p$ for LS-HiPPP is 14.2\%.
At the optimal point, LS-HiPPP can achieve the battery power utilization $\mathcal{U}^{*}_P$ of 95.2\%, however, C-PPP can only achieve $\mathcal{U}^{*}_P$ of 80.5\%.
The optimal cost of 2-BESS using LS-HiPPP is $119.2 c_{\text{h}}$ and C-PPP is $139.2 c_{\text{h}}$.
Therefore, the total cost of C-PPP is 16.8\% more expensive than that of LS-HiPPP.

\paragraph{Cost Analysis Based on Utilization Including Derating}
%We next show an analysis by including the derating. 
Derating and captured value ensure that the power capability of a particular 2-BESS is within a 99.85\,\% confidence level when the batteries are drawn from a statistically distributed supply.
In this case, to build a 2-BESS with 100\,kW (at 1\,C Rate) power output capability, the cost of 2-BESS $c_{_\text{2Bess}}$ is
\begin{align} \label{eqn:total_cost_2}
    c_{_\text{2Bess}} = \frac{100 \,\text{kW}}{\mathcal{C}_v(\mathcal{\hat{R}}_p)}\; c_{_\text{batt}} +
    \frac{100\,\text{kWh}}{\mathcal{C}_v(\mathcal{\hat{R}}_p)}  \; \mathcal{\hat{R}}_p c_{_\text{conv}},
\end{align}
where $\mathcal{C}_v$ is the power captured value of the batteries. The cost of 2-BESS $c_{_\text{2Bess}}$ varies with the cost of power converters $\mathcal{\hat{R}}_p$. The minimal cost is obtained if the marginal cost of power converters equals to the negative of the marginal cost of second-use batteries
\begin{align} \label{eqn:opt_cost_2}
    c_{_\text{batt}} \, \frac{\partial \mathcal{C}_v}{\partial \mathcal{\hat{R}}_p} \Bigg\rvert_{\mathcal{\hat{R}}_p^*} = c_{_\text{conv}} \, \left(\mathcal{U^{*}_P}- \frac{\partial \mathcal{C}_v}{\partial \mathcal{\hat{R}}_p} \Bigg\rvert_{\mathcal{\hat{R}}_p^*} \mathcal{\hat{R}}^{*}_p\right),
\end{align}
where $\mathcal{C}^{*}_v$ is the optimal captured value of the batteries and $\mathcal{\hat{R}}^{*}_p$ is optimal power converter rating.
For example, if the cost of batteries $c_{_\text{batt}}$ and power converters $c_{_\text{conv}}$ are both equal to $c_{\text{h}}$, 
by using Fig.\,\ref{fig:captured_value_reverse} and solving (\ref{eqn:total_cost_2}) and (\ref{eqn:opt_cost_2}), we can compare the optimal cost of different power processing architectures as show in Table\,\ref{table:cost_pp_compare_2}.
\begin{table}[tbp]
    \caption{Cost Comparison of Three Power Processing Architectures Including the Derating}
    \label{table:cost_pp_compare_2}
    \centering
    \begin{tabular}{ccc}
    \toprule
        \textbf{} &\textbf{Batteries (\$)} &\textbf{Converters (\$)} \\
        \midrule
        \textbf{LS-HiPPP} & 106.8$\,c_{\text{h}}$ & 28.7$\,c_{\text{h}}$  \\
        \midrule
        \textbf{C-PPP} & 121.5$\,c_{\text{h}}$ & 54.1$\,c_{\text{h}}$  \\
        \midrule
        \textbf{FPP} & 125 $\,c_{\text{h}}$ & 125 $\,c_{\text{h}}$ \\
    \bottomrule
    \end{tabular}
\end{table}

We simplified the captured value analysis of FPP by assuming that we can access all the capabilities of the batteries. This provides a favorable bound on the derating of FPP.
Given $N$ batteries sampled from a Gaussian battery supply with 20\,\% heterogeneity, the resulting 2-BESS follows a Gaussian distribution with heterogeneity
\begin{align}
    \frac{\sigma_{2b}}{\mu_{2b}} = 20\% \times \frac{1}{\sqrt{N}}.
\end{align}
The $3\sigma$ power derating of FPP is
\begin{align}
    \mathcal{D}_f = 1-20\% \times \frac{1}{\sqrt{N}} \times 3.
\end{align}

The optimal power converter rating $\mathcal{\hat{R}}^{*}_p$ for C-PPP is 44.5\% while $\mathcal{\hat{R}}^{*}_p$ for LS-HiPPP is 26.9\%.
At the optimal point, LS-HiPPP can achieve the battery power utilization $\mathcal{U}^{*}_P$ of 93.6\%, however, C-PPP can only achieve $\mathcal{U}^{*}_P$ of 82.3\%.
The optimal cost of 2-BESS using LS-HiPPP is $135.5 c_{\text{h}}$ and C-PPP is $175.6 c_{\text{h}}$.
Therefore, the total cost of C-PPP is 29.6\% more expensive than that of LS-HiPPP.

In summary, the cost of LS-HiPPP is lower than the other two state-of-the-art power processing architectures.
According to \cite{Neubauer2015}, the cost of second-use batteries is comparable to the cost of power converters, but as batteries become cheaper than power converters,
we need to reduce the amount of power processing for 2-BESS to be competitive.

\section{Conclusion and Future Work}
By utilizing a new hierarchical power processing architecture, LS\nobreakdash-HiPPP achieves better battery utilization and higher system efficiency for the same converter ratings in comparison to FPP and C\nobreakdash-PPP.
By incorporating the heterogeneity statistics of the second-use battery supply, LS-HiPPP is less sensitive to the individual battery variation and has higher power derating as well as power captured value in comparison to FPP and C\nobreakdash-PPP.
At 95\% battery utilization, only one-fifth of the power converter rating is needed for LS\nobreakdash-HiPPP in comparison to FPP.  With the cost of power converters that scale as \$/kW, this corresponds to \mbox{one-fifth} of the cost. Additionally, for the same converter power rating, LS\nobreakdash-HiPPP has the highest efficiency when compared to C\nobreakdash-PPP and FPP.

In the future, we can (1) include more advanced battery modeling and prognostics techniques in the BESS design; (2) use machine learning/AI to extend the strategies described in this paper.

%\appendices

%Appendixes, if needed, appear before the acknowledgment.

%\section*{Acknowledgment}

%\section*{References and Footnotes}

%\section{References}
\bibliographystyle{ieeetr}
\bibliography{library_fixed.bib}

\begin{IEEEbiography}[{\includegraphics[width=1in,height=1.25in,clip,keepaspectratio]{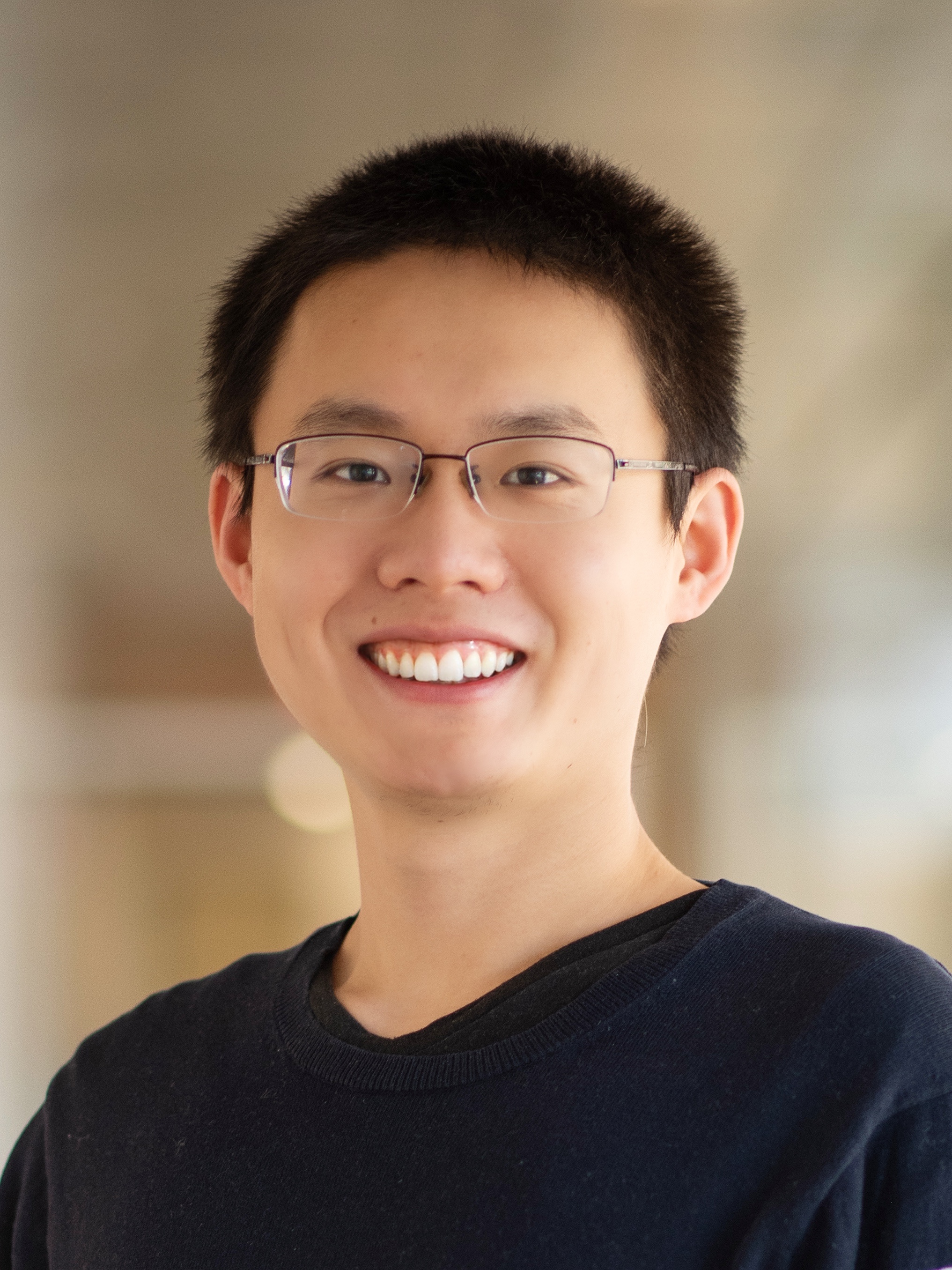}}]{Xiaofan Cui} (Student Member, IEEE) received the B.S. degree in electrical engineering and automation from Tsinghua University, Beijing, China, in 2016. He received the M.S. degree in electrical engineering with the University of Michigan, Ann Arbor, MI, USA, and is currently working toward the Ph.D. degree in electrical and computer engineering.
He was a Visiting Student with Stanford University, Stanford, CA, USA, during the summer of 2015 and a Research Intern with National Renewable Energy Laboratory (NREL) during the summer of 2021.
His research interests include the modeling, control, and design of the energy systems and high-performance power electronics.
Mr. Cui was the recipient of Richard F. and Eleanor A. Towner Prize for Distinguished Academic Achievement at the University of Michigan in 2021.
\end{IEEEbiography}

\begin{IEEEbiography}[{\includegraphics[width=1in,height=1.25in,clip,keepaspectratio]{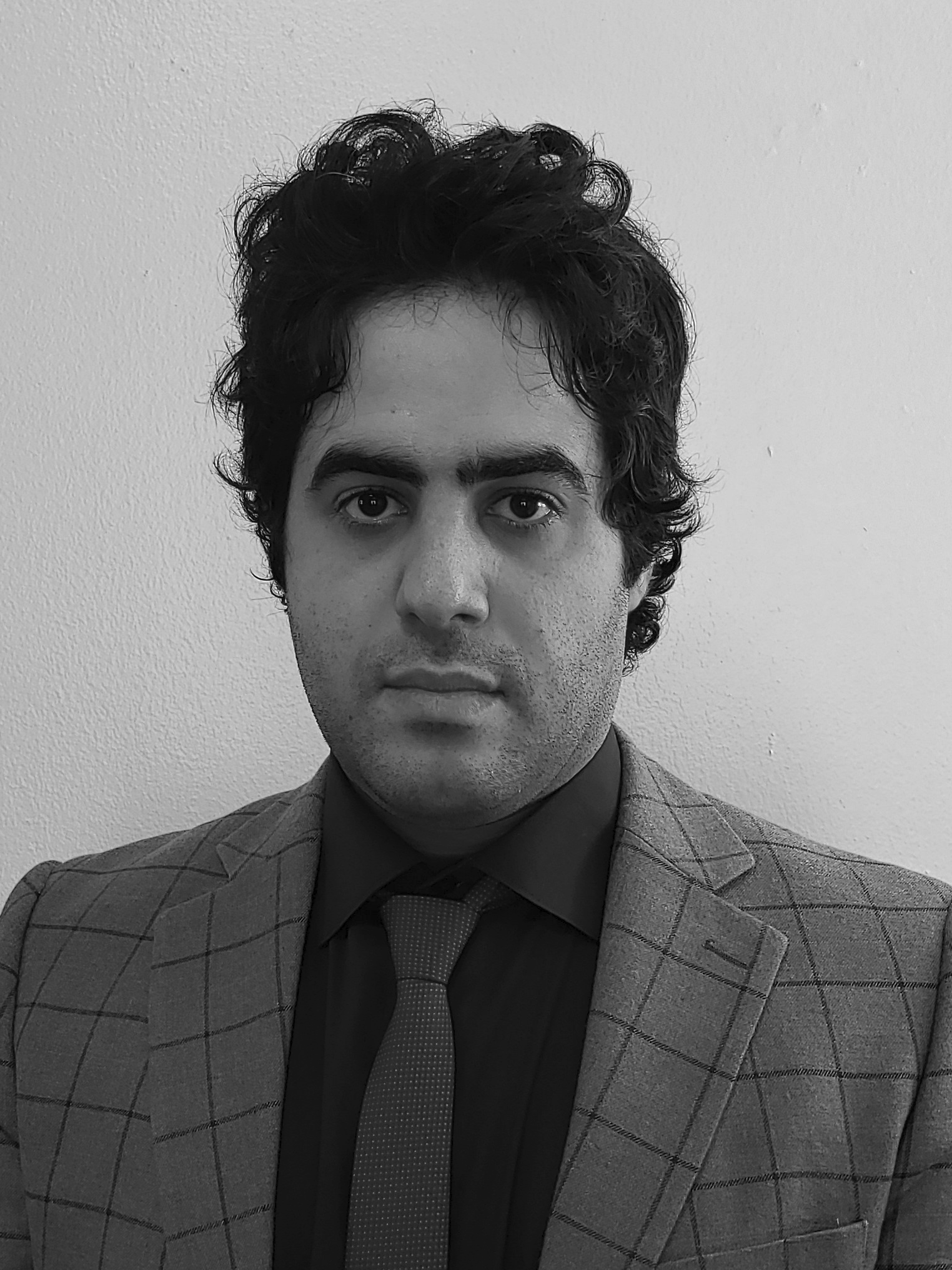}}]{Alireza Ramyar} (Student Member, IEEE) received the B.Sc. degree in electrical engineering from the Sharif University of Technology, Tehran, Iran, in 2013, and the M.Sc. degree in electrical engineering from the University of Tehran, Tehran, Iran, in 2015. He is currently working toward his Ph.D. degree with the Department of Electrical Engineering and Computer Science, University of Michigan, Ann Arbor, MI, USA. His main research interest is designing, modeling, and controlling power electronics for renewable energy and energy storage systems. His complementary interests include circuits and systems for measurement.
\end{IEEEbiography}

\begin{IEEEbiography}[{\includegraphics[width=1in,height=1.25in,clip,keepaspectratio]{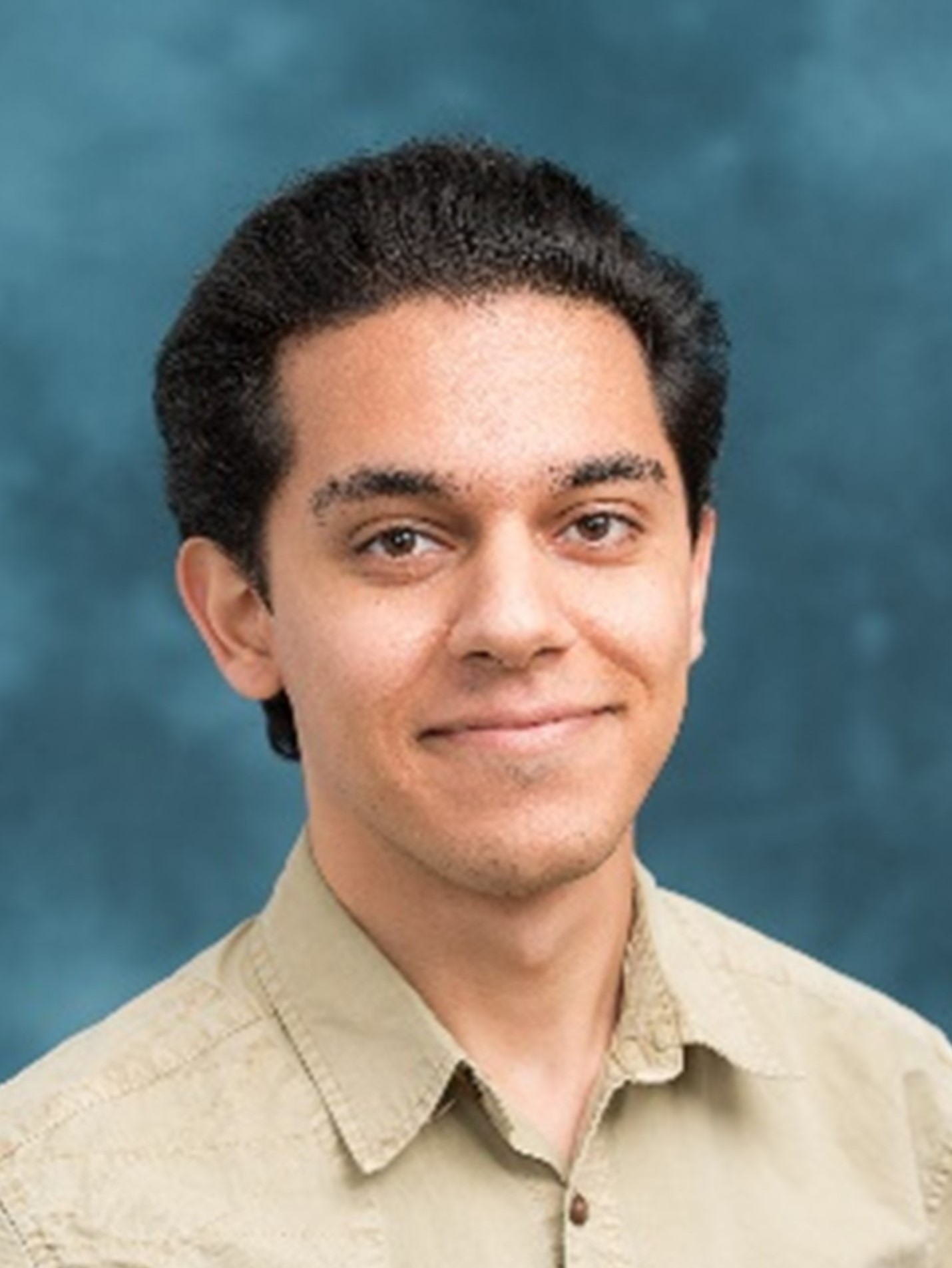}}]{Peyman Mohtat} (Member, IEEE) received the B.S. degree from University of Tehran, Iran, in 2015, and M.S. and the Ph.D. degrees from University of Michigan, Ann Arbor, MI, USA, in 2017 and 2021, respectively, all in mechanical engineering.

His research interests focus on physics-based modeling for lithium-ion batteries and fuel cells. He was involved in projects on battery degradation diagnostics using cell swelling and data-driven models for fuel cells. 
\end{IEEEbiography}

\begin{IEEEbiography}[{\includegraphics[width=1in,height=1.25in,clip,keepaspectratio]{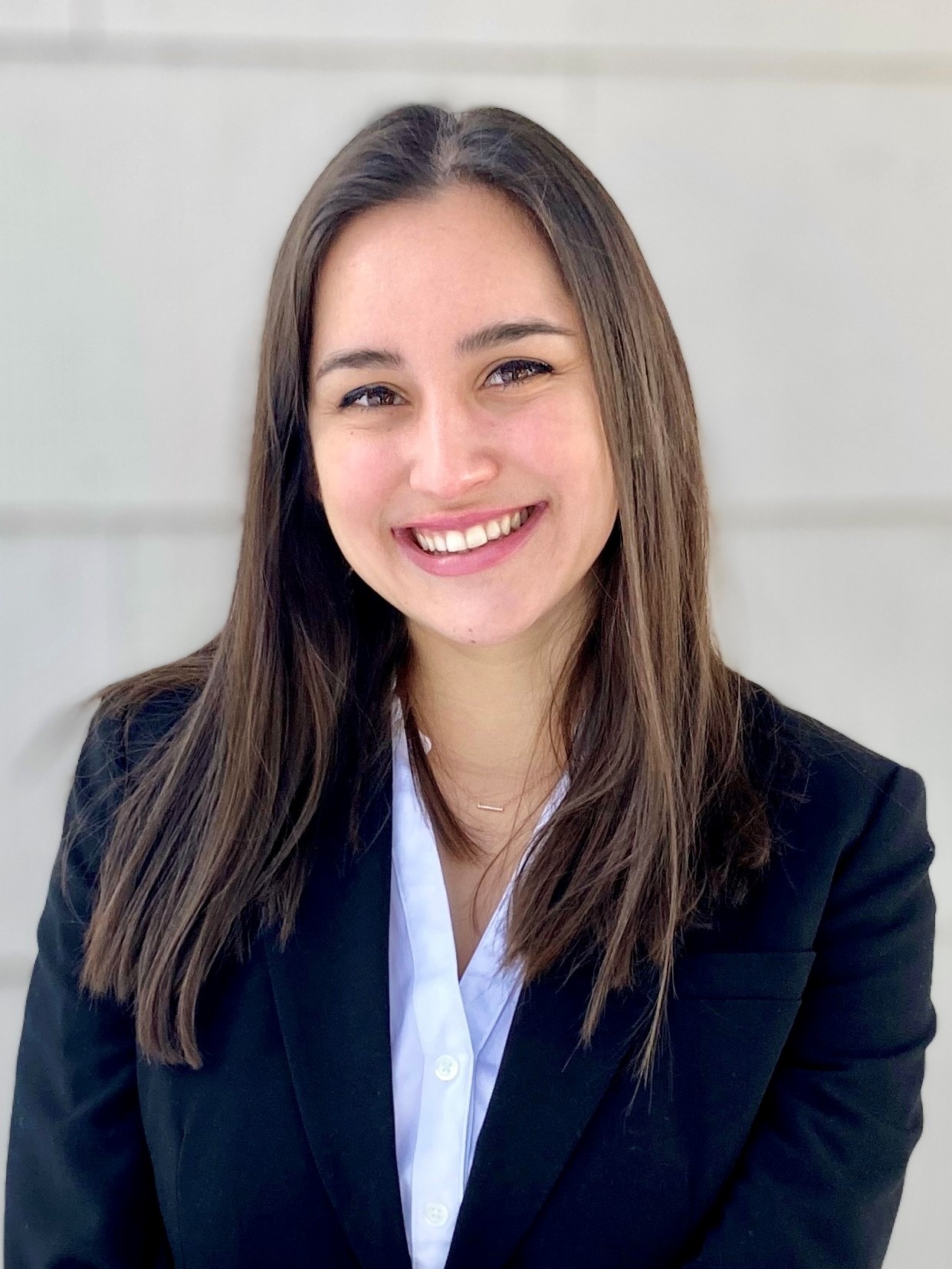}}]{Veronica Contreras} (Student Member, IEEE) received her B.S. degree in Electrical Engineering from the University of California - Davis in 2020. She is currently a GEM Fellow working toward a Ph.D. in Electrical Engineering at the University of Michigan - Ann Arbor.
Veronica was an intern with Lam Research Corporation during the summer of 2020 and 2021. Her research interests include the modeling, control, and design of high-performance power electronics and energy systems.
\end{IEEEbiography}

\begin{IEEEbiography}[{\includegraphics[width=1in,height=1.25in,clip,keepaspectratio]{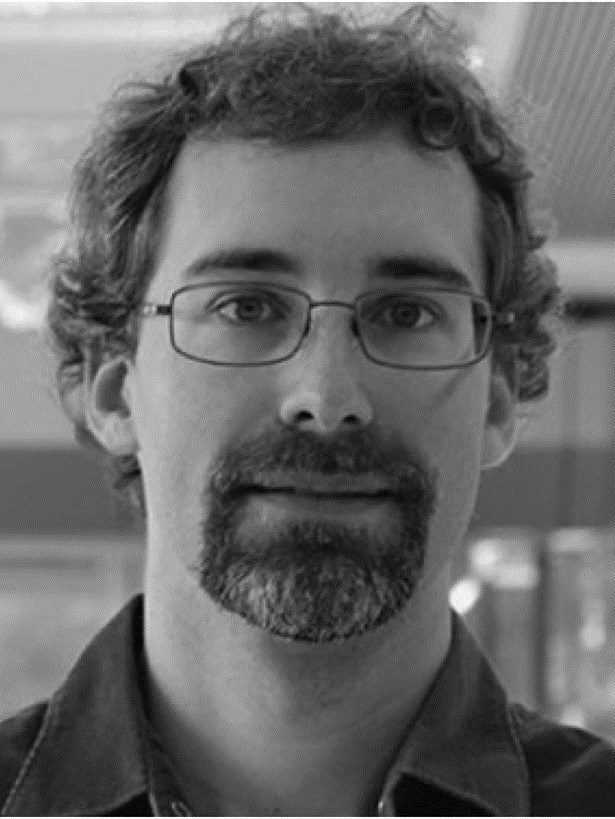}}]{Jason B. Siegel} (Senior Member, IEEE)  received the bachelor’s and the Ph.D. degrees in electrical engineering from the University of Michigan, Ann Arbor, MI, USA, in 2004 and 2010, respectively.

After a two-year postdoc, he joined the faculty as a Research Scientist with the Department of Mechanical Engineering, University of Michigan, in 2012. His research interests include physics-based modeling and control of energy storage and conversion systems including lithium-ion batteries and proton exchange membrane fuel cells. He has coauthored more than 30 journal articles.
\end{IEEEbiography}

\begin{IEEEbiography}[{\includegraphics[width=1in,height=1.25in,clip,keepaspectratio]{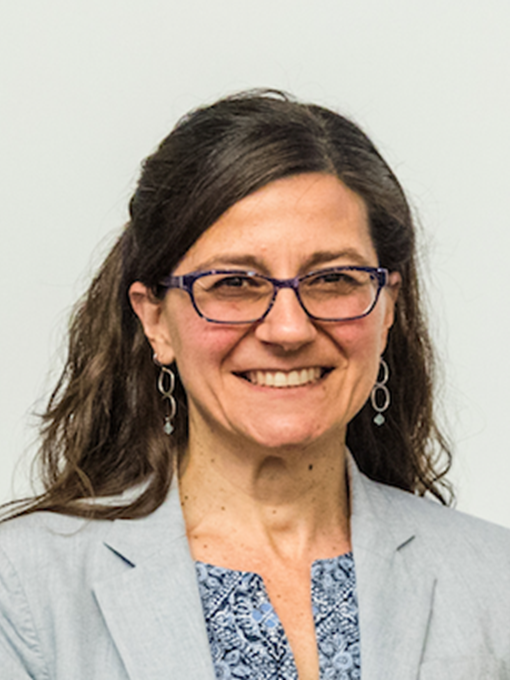}}]{Anna G. Stefanopoulou} (Fellow, IEEE) is the William Clay Ford Professor of Technology at the University of Michigan. She was an assistant professor at the University of California, Santa Barbara, a visiting professor at ETH, Zurich, and a technical specialist at Ford. She earned her diploma in Naval Architecture and Marine Engineering (91, NTUA, Athens), her PhD in Electrical Engineering (96, UMICH, Ann Arbor). She has one book, 21 US patents, 400 publications (7 of which have received awards) on estimation and multivariable control of engines, fuel cells and batteries. She is a Fellow of the ASME (08), IEEE (09), and SAE (18). 
\end{IEEEbiography}

\begin{IEEEbiography}[{\includegraphics[width=1in,height=1.25in,clip,keepaspectratio]{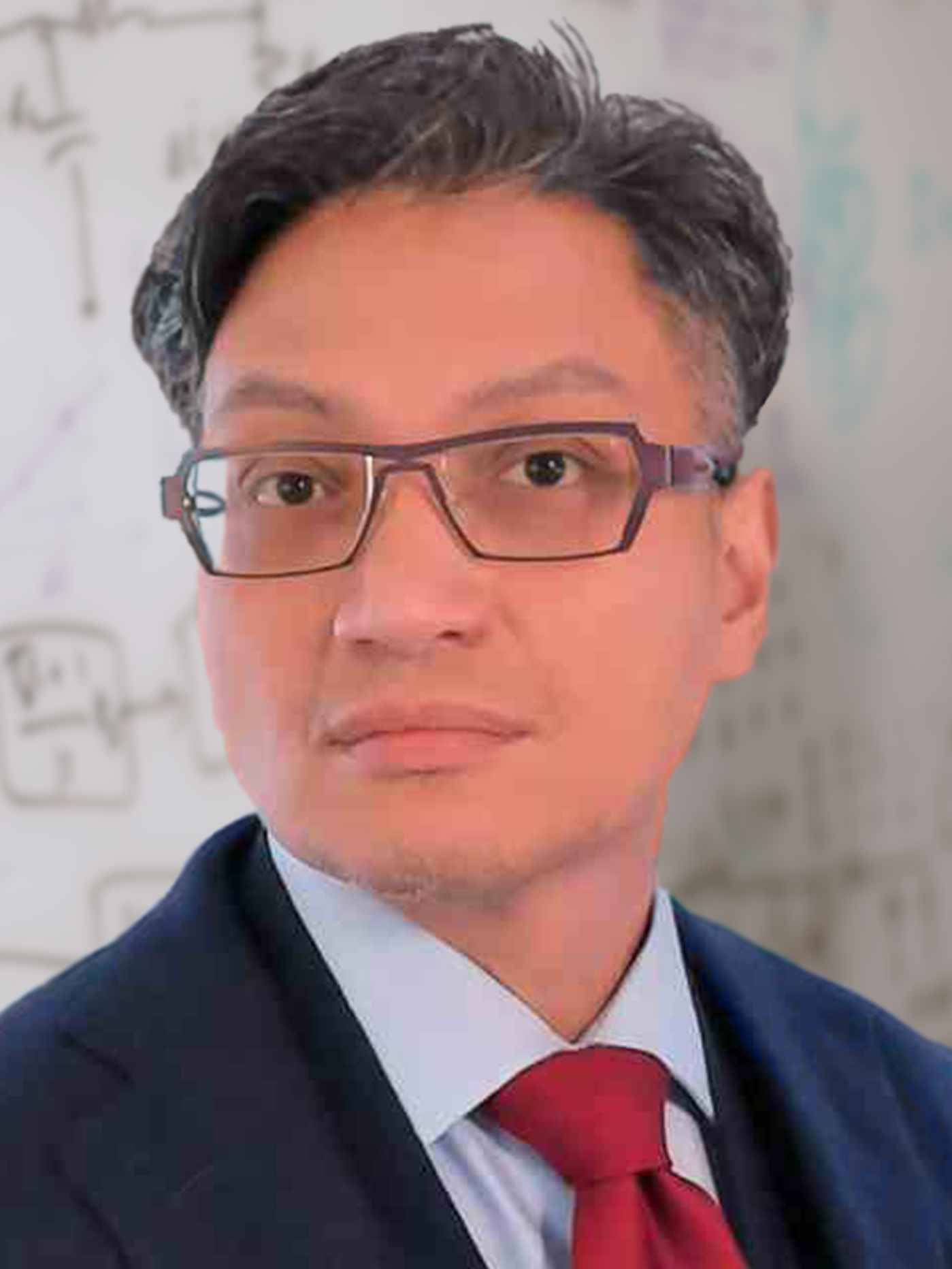}}]{Al-Thaddeus Avestruz} (Senior Member, IEEE) received the SB in Physics with Electrical Engineering, the SM and EE degrees in 2006, and the PhD in 2016 in Electrical Engineering and Computer Science from the Massachusetts Institute of Technology, Cambridge, MA, USA.  He is currently an Assistant Professor in Electrical and Computer Engineering at the University of Michigan, Ann Arbor, MI, USA.  His research interests include the design, modeling, and control of high-performance power electronics and wireless power transfer for energy, mobility, medicine, and the Internet of Things.  He has complementary interests in circuits and systems for sensing, electromagnetic systems, feedback and controls, renewable energy, automotive, biomedical, and consumer applications.  Dr. Avestruz is the chair of TC1: Control and Modeling of Power Electronics for the IEEE Power Electronics Society.  He was the recipient of the IEEE PELS Best ECCE Paper on Emerging Technology Award Oral Presentation in 2019.  He is an Associate Editor of the IEEE Open Journal of Power Electronics, a Guest Associate Editor of the IEEE Journal of Emerging and Selected Topics in Power Electronics, an Associate Technical Program Chair for the IEEE Energy Conversion Congress and Exposition in 2019, a Technical Program Co-Chair for the 2021 IEEE Wireless Power Week, General Chair for the 24th IEEE Workshop on Control and Modeling for Power Electronics in 2023.  
He received the NSF CAREER award in 2022.
He has over a decade of industry and entrepreneurial experience, and holds 11 issued U.S. patents.
\end{IEEEbiography}

\EOD

\end{document}